% !TEX encoding = UTF-8 Unicode
% !TEX TS-program = pdftex
% !TEX spellcheck = en-EN

%\input diagrams.tex
 \input pictex.tex   %% Dovrebbe essere in path (altrimenti vedi in Doc)

% $Id: dcpic.sty,v 1.24 2002/11/25 13:51:57 pedro Exp $
%% DC-PiCTeX
%% Realizado por Pedro Quaresma de Almeida, Coimbra 
%% 11/1990 (vers{\~a}o 1.0); 10/1991 (vers{\~a}o 1.1);
%%  9/1993 (vers{\~a}o 1.2);  3/1995 (vers{\~a}o 1.3);
%%  7/1996 (vers{\~a}o 2.1);
%%  5/2001 (vers{\~a}o 3.0); 11/2001 (vers{\~a}o 3.1);
%%  1/2002 (vers{\~a}o 3.2)
%%  5/2002 (vers?o 4.0)
\immediate\write10{Package DCpic 2002/05/16 v4.0}

\catcode`!=11 %  ***** THIS MUST NEVER BE OMITTED (Ver PiCTeX)

\newcount\aux%
\newcount\auxa%
\newcount\auxb%
\newcount\m%
\newcount\n%
\newcount\x%
\newcount\y%
\newcount\xl%
\newcount\yl%
\newcount\d%
\newcount\dnm%
\newcount\xa%
\newcount\xb%
\newcount\xmed%
\newcount\xc%
\newcount\xd%
\newcount\ya%
\newcount\yb%
\newcount\ymed%
\newcount\yc%
\newcount\yd
%% "vari·veis globais"
\newcount\expansao%
\newcount\tipografo%       vers?o 4.0
\newcount\distanciaobjmor% vers?o 4.0
\newcount\tipoarco%        vers?o 4.0
%\newif\ifarredondada%        vers?o 4.0 (valor inicial "false")
\newif\ifpara%
%% version 3.2
\newbox\caixa%
\newbox\caixaaux%
\newif\ifnvazia%
\newif\ifvazia%
\newif\ifcompara%
\newif\ifdiferentes%
\newcount\xaux%
\newcount\yaux%
\newcount\guardaauxa%
\newcount\alt%
\newcount\larg%
\newcount\prof%
%% para os ajustes
\newcount\auxqx
\newcount\auxqy
\newif\ifajusta%
\newif\ifajustadist
\def\objPartida{}%
\def\objChegada{}%
\def\objNulo{}%

%% 
%% Stack specification
%%

%%
%% Emtpy stack
%%
\def\!vazia{:}

%%
%% Is Empty? : Stack -> Bool
%%
%% nvazia - True if Not Empy
%% vazia  - True if Empty
\def\!pilhanvazia#1{\let\arg=#1%
\if:\arg\ \nvaziafalse\vaziatrue \else \nvaziatrue\vaziafalse\fi}

%%
%% Push : Elems x Stack -> Stack
%%
\def\!coloca#1#2{\edef\pilha{#1.#2}}

%%
%% Top : Stack -> Elems
%%
%% the empty stack is not taken care
%% the element is "kept" ("guardado") 
\def\!guarda(#1)(#2,#3)(#4,#5,#6){\def\id{#1}%
\xaux=#2%
\yaux=#3%
\alt=#4%
\larg=#5%
\prof=#6%
}

\def\!topaux#1.#2:{\!guarda#1}
\def\!topo#1{\expandafter\!topaux#1}

%%
%% Pop : Stack -> Stack
%%
%% the empty stack is not taken care
\def\!popaux#1.#2:{\def\pilha{#2:}}
\def\!retira#1{\expandafter\!popaux#1}

%%
%% Compares words : Word x Word -> Bool
%%
%% compara - True if equal
%% diferentes - True if not equal
\def\!comparaaux#1#2{\let\argA=#1\let\argB=#2%
\ifx\argA\argB\comparatrue\diferentesfalse\else\comparafalse\diferentestrue\fi}

\def\!compara#1#2{\!comparaaux{#1}{#2}}

%%Comando Interno
%% Valor absoluto (absolute value)
%% \absoluto{n}{absn}
%% entrada
%%  n - natural
%% sa{\'\i}da
%%  absn - o valor absoluto de n
\def\!absoluto#1#2{\n=#1%
  \ifnum \n > 0
    #2=\n
  \else
    \multiply \n by -1
    #2=\n
  \fi}

%% Name definitions for edge types and directions
\def\solidarrow{0}

%% Name definitions for edge label placement
\def\atright{-1}
\def\atleft{1}
%% Tip direction for curved edges
\def\pup{0}

\def\pright{2}
\def\pleft{3}
%% Type of graph
\def\commdiag{0}

%% Posicionamento da etiquetas nos grafos

%%Comando Interno
%% Ajusta a dist{\^a}ncia entre as setas e os objectos em fun{\c c}{\~a}o das
%% dimens{\~o}es destes {\'u}ltimos
%% \ajusta{x}{xl}{y}{yl}{d}{Objecto}
%% entrada
%%  (x,y) e (xl,yl), coordenadas dos pontos de {\'\i}nicio e fim da seta
%%  d, dist{\^a}ncia especificada pelo utilizador ou 10 (valor por
%%  omiss{\~a}o), Objecto d{\'a}-nos a refer{\^e}ncia do objecto ao qual se est{\'a} a
%%  efectuar o ajuste.
%% sa{\'\i}da
%%  d, dist{\^a}ncia alterada.
%% 
%% A dist{\^a}ncia alterada {\'e} o maior valor entre 10 e as dimens{\~o}es
%% apropriadas da caixa que cont{\^e}m o objecto. 
%% Se o utilizador especificar um valor essa especifica{\c c}{\~a}o
%% n{\~a}o {\'e} alterada.
%%
%% Se a seta {\'e} horizontal usa-se o valor da largura
%% Se a seta {\'e} vertical usa-se:
%%  o valor da altura se a seta est{\'a} no 1o ou 2o quadrante
%%  o valor da profundidade se a seta est{\'a} no 3o ou 4o quadrante
%% Se a seta {\'e} {\'o}bliqua vai-se escolher o valor conforme:
%%  de 315 a  45 graus usa-se a largura
%%  de  45 a 135 graus usa-se a altura
%%  de 135 a 225 graus usa-se a largura
%%  de 225 a 315 graus usa-se a profundidade
\def\!ajusta#1#2#3#4#5#6{\aux=#5%
  \let\auxobj=#6%
  \ifcase \tipografo    % diagramas comutativos
    \ifnum\number\aux=10 
      \ajustadisttrue % se o valor È o valor por omiss?o ajusta
    \else
      \ajustadistfalse  % caso contr·rio n?o ajusta
    \fi
  \else  % grafos (dirigidos, n?o dirigidos, com molduras)
   \ajustadistfalse
%  \or  % grafos n?o dirigidos
%   \ajustadistfalse
%  \else % grafos dirigidos com molduras circulares nos objectos
%    \ifnum\number\aux=8 
%      \ajustadisttrue  % se o valor È o valor por omiss?o ajusta
%    \else
%      \ajustadistfalse % caso contr·rio n?o ajusta
%    \fi
  \fi
  \ifajustadist
%  \tiny Vou ajustar %%%%%%%%%%%%%%%%%%%%%%%%%%%%%%
%  \ifnum\number\aux=10% verificar se s{\~a}o os valores por omiss{\~a}o
   \let\pilhaaux=\pilha%
   \loop%
     \!topo{\pilha}%
     \!retira{\pilha}%
     \!compara{\id}{\auxobj}%
     \ifcompara\nvaziafalse \else\!pilhanvazia\pilha \fi%
     \ifnvazia%
   \repeat%
%% rep{\~o}e os valores na pilha
   \let\pilha=\pilhaaux%
   \ifvazia%
    \ifdiferentes%
%%
%% N{\~a}o {\'e} poss{\'\i}vel efectuar o ajuste dado o utilizador n{\~a}o ter
%% especificado uma etiqueta para o objecto em quest{\~a}o. {\'E} dado o
%% valor de 10, igual ao valor por omiss{\~a}o.
%%
     \larg=1310720% n{\~a}o faz o ajuste
     \prof=655360%
     \alt=655360%
    \fi%
   \fi%
   \divide\larg by 131072
   \divide\prof by 65536
   \divide\alt by 65536
   \ifnum\number\y=\number\yl
%% Caso 1 -- seta horizontal
%%
%% divide-se por 131072 para se obter metade da largura da caixa em
%% pontos (pt), isto dado que o texto est{\'a} centrado na caixa. Soma-se
%% mais tr{\^e}s, que constitue um ajuste imp{\'\i}rico.
    \advance\larg by 3
    \ifnum\number\larg>\aux
     #5=\larg
    \fi
   \else
    \ifnum\number\x=\number\xl
     \ifnum\number\yl>\number\y
%% Caso 2.1 -- seta vertical de cima para baixa
%%
      \ifnum\number\alt>\aux
       #5=\alt
      \fi
     \else
%% Caso 2.2 -- seta vertical de baixo para cima
%%
%% divide-se por 65536 para se obter a altura da caixa em pt. O ajuste
%% de 5 foi obtido imp{\'\i}ricamente
      \advance\prof by 5
      \ifnum\number\prof>\aux
       #5=\prof
      \fi
     \fi
    \else
%% Caso 3 -- seta obl{\'\i}qua 
%% Caso 3.1 de 315o a  45o; |x-xl|>|y-yl| e
%% Caso 3.3 de 135o a 225o; |x-xl|>|y-yl|; Largura
     \auxqx=\x
     \advance\auxqx by -\xl
     \!absoluto{\auxqx}{\auxqx}%
     \auxqy=\y
     \advance\auxqy by -\yl
     \!absoluto{\auxqy}{\auxqy}%
     \ifnum\auxqx>\auxqy
      \ifnum\larg<10
       \larg=10
      \fi
      \advance\larg by 3
      #5=\larg
     \else
%% Caso 3.2 de  45o a 135o; |x-xl|<|y-yl| e y>0; Largura
      \ifnum\yl>\y
       \ifnum\larg<10
        \larg=10
       \fi
      \advance\alt by 6
       #5=\alt
      \else
%% Caso 3.4 de 225o a 315o; |x-xl|<|y-yl| e y<0; Profundidade
      \advance\prof by 11
       #5=\prof
      \fi
     \fi
    \fi
   \fi
\fi} % o ramo "else" {\'e} omisso

%%Comando Interno
%% C{\'a}lculo da Raiz Quadrada
%% raiz{n}{m}
%% entrada
%%   n - natural
%% sa{\'\i}da
%%   n - natural
%%   m - maior natural contido na raiz quadrada de n
\def\!raiz#1#2{\n=#1%
  \m=1%
  \loop
    \aux=\m%
    \advance \aux by 1%
    \multiply \aux by \aux%
    \ifnum \aux < \n%
      \advance \m by 1%
      \paratrue%
    \else\ifnum \aux=\n%
      \advance \m by 1%
      \paratrue%
       \else\parafalse%
       \fi
    \fi
  \ifpara%
  \repeat
#2=\m}

%%Comando Interno
%% Calcula os pontos de 
%%       come{\c c}o da "seta"
%%       fim da "seta"
%%   coloca{\c c}{\~a}o do s{\'\i}mbolo
%% 
%% ucoord{x1}{x2}{x3}{x4}{x5}{x6}{+|- 1}
%% entrada
%%   x1,x2,x3,x4,x5
%% sa{\'\i}da
%%   x6
%%  
%%              x2 - x1
%%  x6 = x3 +|- ------- x4
%%                 x5
\def\!ucoord#1#2#3#4#5#6#7{\aux=#2%
  \advance \aux by -#1%
  \multiply \aux by #4%
  \divide \aux by #5%
  \ifnum #7 = -1 \multiply \aux by -1 \fi%
  \advance \aux by #3%
#6=\aux}

%%Comando Interno 
%% C{\'a}lculo do Quadrado da Dist{\^a}ncia Euclidiana entre dois pontos 
%% quadrado{n}{m}{l}
%% entrada
%%   n - natural
%%   m - natural
%% sa{\'\i}da
%%   l = (n-m)*(n-m)
\def\!quadrado#1#2#3{\aux=#1%
  \advance \aux by -#2%
  \multiply \aux by \aux%
#3=\aux}

%%Comando Interno
%% C{\'a}lculo auxiliar para determinar a dist{\^a}ncia entre o nome do
%% morfismo e a seta.
%% entrada
%%     (x,y), (x',y') e o nome do morfismo
%% sa{\'\i}da
%%     dnm - dist{\^a}ncia do nome ao morfismo respectivo devidamente
%%     compensada pelo tamanho do objecto
%% Observa{\c c}{\~o}es
%%     A compensa{\c c}{\~a}o s{\'o} est{\'a} a ser feita para setas
%%     horizontais e verticais. As obl{\'\i}quas s{\~a}o tratadas de
%%     outra forma.
%% algoritmo
%%  caixa0 <- nome do morfismo
%%  se x-xl = 0 entao                   {recta vertical}
%%     aux <- largura da caixa0
%%     dnm <- convers{\~a}o-sp-pt(aux)/2+3
%%  sen{\~a}o                               {recta n{\~a}o vertical}
%%     se y-yl = 0 entao                {recta horizontal}
%%        aux <- altura+profundidade da caixa0
%%        dnm <- convers{\~a}o-sp-pt(aux)/2+3
%%     sen{\~a}o                            {recta obl{\'\i}qua}
%%        dnm <- 3
%%     fimse
%%  fimse
%% fimalgoritmo
\def\!distnomemor#1#2#3#4#5#6{\setbox0=\hbox{#5}%
  \aux=#1
  \advance \aux by -#3
  \ifnum \aux=0
     \aux=\wd0 \divide \aux by 131072
     \advance \aux by 3
     #6=\aux
  \else
     \aux=#2
     \advance \aux by -#4
     \ifnum \aux=0
        \aux=\ht0 \advance \aux by \dp0 \divide \aux by 131072
        \advance \aux by 3
        #6=\aux%
     \else
     #6=3
     \fi
   \fi
}

%%
%% O ambiente "begindc...enddc"
%%
\def\begindc#1{\!ifnextchar[{\!begindc{#1}}{\!begindc{#1}[30]}}
\def\!begindc#1[#2]{\beginpicture 
  \let\pilha=\!vazia
  \setcoordinatesystem units <1pt,1pt>
  \expansao=#2
  \ifcase #1
    \distanciaobjmor=10
    \tipoarco=0         % seta
    \tipografo=0        % diagrama comutativo
  \or
    \distanciaobjmor=2
    \tipoarco=0         % seta 
    \tipografo=1        % grafo orientado
  \or
    \distanciaobjmor=1
    \tipoarco=2         % linha
    \tipografo=2        % grafo n?o orientado
  \or
    \distanciaobjmor=8
    \tipoarco=0         % seta 
    \tipografo=3        % grafo orientado
%    \arredondadotrue    % objectos com molduras circulares
  \or
    \distanciaobjmor=8
    \tipoarco=2         % linha
    \tipografo=4        % grafo n?o orientado
%    \arredondadotrue    % objectos com molduras circulares
  \fi}

\def\enddc{\endpicture}

%%
%% Comando para construir a "seta" entre dois objectos
%%
%% Os pontos definidores da seta e da etiqueta respectiva s{\~a}o:
%% 
%%                (xd,yd)
%%                   o
%%                   |
%%  o------o---------o---------o------o
%%(x,y) (xa,ya)   (xm,ym)   (xb,yb)(xl,yl)
%%
\def\mor{%
  \!ifnextchar({\!morxy}{\!morObjA}}
\def\!morxy(#1,#2){%
  \!ifnextchar({\!morxyl{#1}{#2}}{\!morObjB{#1}{#2}}}
\def\!morxyl#1#2(#3,#4){%
  \!ifnextchar[{\!mora{#1}{#2}{#3}{#4}}{\!mora{#1}{#2}{#3}{#4}[\number\distanciaobjmor,\number\distanciaobjmor]}}%
\def\!morObjA#1{%
 \let\pilhaaux=\pilha%
 \def\objPartida{#1}%
 \loop%
    \!topo\pilha%
    \!retira\pilha%
    \!compara{\id}{\objPartida}%
    \ifcompara \nvaziafalse \else \!pilhanvazia\pilha \fi%
   \ifnvazia%
 \repeat%
 \ifvazia%
  \ifdiferentes%
%%
%% Mensagem de erro e atribui{\c c}{\~a}o de valores fict{\'\i}cios aos 
%% argumentos dos comandos que se seguem.
%%
   Error: Incorrect label specification%
   \xaux=1%
   \yaux=1%
  \fi%
 \fi% 
 \let\pilha=\pilhaaux%
 \!ifnextchar({\!morxyl{\number\xaux}{\number\yaux}}{\!morObjB{\number\xaux}{\number\yaux}}}
\def\!morObjB#1#2#3{%
  \x=#1
  \y=#2
 \def\objChegada{#3}%
 \let\pilhaaux=\pilha%
 \loop
    \!topo\pilha %
    \!retira\pilha%
    \!compara{\id}{\objChegada}%
    \ifcompara \nvaziafalse \else \!pilhanvazia\pilha \fi
   \ifnvazia
 \repeat
 \ifvazia
  \ifdiferentes%
%%
%% Mensagem de erro e atribui{\c c}{\~a}o de valores fict{\'\i}cios aos 
%% argumentos dos comandos que se seguem.
%%
   Error: Incorrect label specification
   \xaux=\x%
   \advance\xaux by \x%
   \yaux=\y%
   \advance\yaux by \y%
  \fi
 \fi
 \let\pilha=\pilhaaux
 \!ifnextchar[{\!mora{\number\x}{\number\y}{\number\xaux}{\number\yaux}}{\!mora{\number\x}{\number\y}{\number\xaux}{\number\yaux}[\number\distanciaobjmor,\number\distanciaobjmor]}}
\def\!mora#1#2#3#4[#5,#6]#7{%
  \!ifnextchar[{\!morb{#1}{#2}{#3}{#4}{#5}{#6}{#7}}{\!morb{#1}{#2}{#3}{#4}{#5}{#6}{#7}[1,\number\tipoarco] }}
\def\!morb#1#2#3#4#5#6#7[#8,#9]{\x=#1%
  \y=#2%
  \xl=#3%
  \yl=#4%
  \multiply \x by \expansao%
  \multiply \y by \expansao%
  \multiply \xl by \expansao%
  \multiply \yl by \expansao%
%%
%% calcular a dist{\^a}ncia Euclidiana entre dois pontos
%% d = \sqrt((x-xl)^2+(y-yl)^2)
%%
  \!quadrado{\number\x}{\number\xl}{\auxa}%
  \!quadrado{\number\y}{\number\yl}{\auxb}%
  \d=\auxa%
  \advance \d by \auxb%
  \!raiz{\d}{\d}%
%%
%% o ponto (xa,ya) est{\'a} {\`a} dist{\^a}ncia #5 (valor por omiss{\~a}o 10) do ponto
%% (x,y)
%%
%% como existem dois pontos em considera{\c c}{\~a}o, o ponto de partida e o
%% ponto de chegada, vai sei necess{\'a}rio recuperar de novo os seus
%% valores por pesquisa na pilha
  \auxa=#5
  \!compara{\objNulo}{\objPartida}%
  \ifdiferentes% S{\'o} vai fazer o ajuste quando {\'e} necess{\'a}rio
   \!ajusta{\x}{\xl}{\y}{\yl}{\auxa}{\objPartida}%
   \ajustatrue
   \def\objPartida{}% re-inicializar o valor do Objecto de Partida
  \fi
%% vai guardar o valor de auxa (ap{\'o}s ajuste) para ser usado no caso
%% dos morfismos de injec{\c c}{\~a}o.
  \guardaauxa=\auxa
  \!ucoord{\number\x}{\number\xl}{\number\x}{\auxa}{\number\d}{\xa}{1}%
  \!ucoord{\number\y}{\number\yl}{\number\y}{\auxa}{\number\d}{\ya}{1}%
%% auxa vai ter o valor da dist{\^a}ncia entre os objectos menos a
%% dist{\^a}ncia da seta ao objecto (10 por omiss{\~a}o)
  \auxa=\d%
%%
%% o ponto (xb,yb) est{\'a} {\`a} dist{\^a}ncia #6 (valor por omiss{\~a}o 10) do ponto
%% (xl,yl)
%%
  \auxb=#6
  \!compara{\objNulo}{\objChegada}%
  \ifdiferentes% S{\'o} vai fazer o ajuste quando {\'e} necess{\'a}rio
%   Vou ajustar
   \!ajusta{\x}{\xl}{\y}{\yl}{\auxb}{\objChegada}%
   \def\objChegada{}% re-inicializar o valor do Objecto de Chegada
  \fi
  \advance \auxa by -\auxb%
  \!ucoord{\number\x}{\number\xl}{\number\x}{\number\auxa}{\number\d}{\xb}{1}%
  \!ucoord{\number\y}{\number\yl}{\number\y}{\number\auxa}{\number\d}{\yb}{1}%
  \xmed=\xa%
  \advance \xmed by \xb%
  \divide \xmed by 2
  \ymed=\ya%
  \advance \ymed by \yb%
  \divide \ymed by 2
  \!distnomemor{\number\x}{\number\y}{\number\xl}{\number\yl}{#7}{\dnm}%
  \!ucoord{\number\y}{\number\yl}{\number\xmed}{\number\dnm}{\number\d}{\xc}{-#8}% 
  \!ucoord{\number\x}{\number\xl}{\number\ymed}{\number\dnm}{\number\d}{\yc}{#8}%
\ifcase #9  % seta s{\'o}lida
  \arrow <4pt> [.2,1.1] from {\xa} {\ya} to {\xb} {\yb}
\or  % seta a tracejado
  \setdashes
  \arrow <4pt> [.2,1.1] from {\xa} {\ya} to {\xb} {\yb}
  \setsolid
\or  % linha s{\'o}lida
  \setlinear
  \plot {\xa} {\ya}  {\xb} {\yb} /
\or  % seta de injec{\c c}{\~a}o
%% C{\'a}lculos auxiliares
%%
%% 3 valor para o raio do "rabo" da "seta"
%%
%% repor o valor de auxa
  \auxa=\guardaauxa
%% dar a compensa{\c c}{\~a}o para o "rabo"
  \advance \auxa by 3%
%%
%% IMPORTANTE os valores de xa e ya v{\~a}o ser alterados
%%
 \!ucoord{\number\x}{\number\xl}{\number\x}{\number\auxa}{\number\d}{\xa}{1}%
 \!ucoord{\number\y}{\number\yl}{\number\y}{\number\auxa}{\number\d}{\ya}{1}%
 \!ucoord{\number\y}{\number\yl}{\number\xa}{3}{\number\d}{\xd}{-1}%
 \!ucoord{\number\x}{\number\xl}{\number\ya}{3}{\number\d}{\yd}{1}%
%% Constru{\c c}{\~a}o da "seta"
  \arrow <4pt> [.2,1.1] from {\xa} {\ya} to {\xb} {\yb}
%% e do seu "rabo"
  \circulararc -180 degrees from {\xa} {\ya} center at {\xd} {\yd}
\or  % seta de aplica{\c c}{\~a}o ("|-->")
  \auxa=3% valor para o meio-segmento do "rabo" da "seta"
%% c{\'a}lculo dos pontos (xmed,ymed) e (xd,yd) para o segmento de recta que
%% define o "rabo" da seta
 \!ucoord{\number\y}{\number\yl}{\number\xa}{\number\auxa}{\number\d}{\xmed}{-1}%
 \!ucoord{\number\x}{\number\xl}{\number\ya}{\number\auxa}{\number\d}{\ymed}{1}%
 \!ucoord{\number\y}{\number\yl}{\number\xa}{\number\auxa}{\number\d}{\xd}{1}%
 \!ucoord{\number\x}{\number\xl}{\number\ya}{\number\auxa}{\number\d}{\yd}{-1}%
%% Constru{\c c}{\~a}o da "seta"
  \arrow <4pt> [.2,1.1] from {\xa} {\ya} to {\xb} {\yb}
%% e do seu "rabo"
  \setlinear
  \plot {\xmed} {\ymed}  {\xd} {\yd} /
\fi
%% Coloca{\c c}{\~a}o do nome do morfismo.
%% Se os morfismos s{\~a}o horizontais ou verticais constro{\'\i} uma caixa
%% centrada no ponto pr{\'e}viamente calculado. Se as setas s{\~a}o
%% obl{\'\i}quas coloca a caixa de forma a n{\~a}o colidir com o morfismo 
%% tendo em aten{\c c}{\~a}o o quadrante assim como a posi{\c c}{\~a}o
%% relativa do morfismo e do respectivo nome.
\auxa=\xl
\advance \auxa by -\x%
\ifnum \auxa=0 
  \put {#7} at {\xc} {\yc}
\else
  \auxb=\yl
  \advance \auxb by -\y%
  \ifnum \auxb=0 \put {#7} at {\xc} {\yc}
  \else 
    \ifnum \auxa > 0 
      \ifnum \auxb > 0
        \ifnum #8=1
          \put {#7} [rb] at {\xc} {\yc}
        \else 
          \put {#7} [lt] at {\xc} {\yc}
        \fi
      \else
        \ifnum #8=1
          \put {#7} [lb] at {\xc} {\yc}
        \else 
          \put {#7} [rt] at {\xc} {\yc}
        \fi
      \fi
    \else
      \ifnum \auxb > 0 
        \ifnum #8=1
          \put {#7} [rt] at {\xc} {\yc}
        \else 
          \put {#7} [lb] at {\xc} {\yc}
        \fi
      \else
        \ifnum #8=1
          \put {#7} [lt] at {\xc} {\yc}
        \else 
          \put {#7} [rb] at {\xc} {\yc}
        \fi
      \fi
    \fi
  \fi
\fi
}

%%
%% Comando para construir a "seta" curvilinea entre dois objectos
%%
%% \cmor(<lista de pontos (n. impar)>){<etiqueta>}
%%
%% Em primeiro lugar {\'e} necess{\'a}rio modificar o comando plot de forma a
%% que a sintaxe de utiliza{\c c}{\~a}o do novo comando seja coerente com a
%% sintaxe dos restantes comandos
%%
\def\modifplot(#1{\!modifqcurve #1}
\def\!modifqcurve(#1,#2){\x=#1%
  \y=#2%
  \multiply \x by \expansao%
  \multiply \y by \expansao%
  \!start (\x,\y)
  \!modifQjoin}
\def\!modifQjoin(#1,#2)(#3,#4){\x=#1%
  \y=#2%
  \xl=#3%
  \yl=#4%
  \multiply \x by \expansao%
  \multiply \y by \expansao%
  \multiply \xl by \expansao%
  \multiply \yl by \expansao%
  \!qjoin (\x,\y) (\xl,\yl)             % \!qjoin  is defined in QUADRATIC
  \!ifnextchar){\!fim}{\!modifQjoin}}
\def\!fim){\ignorespaces}

%%
%% O comando para desenhar a seta vai receber a lista de pontos da qual
%% retira o {\'u}ltimo par de pontos, dependente da escolha dada pelo
%% utilizador a seta vai ser desenhada para cima, para baixo, para a
%% direita ou para a esquerda
%%
\def\setaxy(#1{\!pontosxy #1}
\def\!pontosxy(#1,#2){%
  \!maispontosxy}
\def\!maispontosxy(#1,#2)(#3,#4){%
  \!ifnextchar){\!fimxy#3,#4}{\!maispontosxy}}
\def\!fimxy#1,#2){\x=#1%
  \y=#2
  \multiply \x by \expansao
  \multiply \y by \expansao
  \xl=\x%
  \yl=\y%
  \aux=1%
  \multiply \aux by \auxa%
  \advance\xl by \aux%
  \aux=1%
  \multiply \aux by \auxb%
  \advance\yl by \aux%
  \arrow <4pt> [.2,1.1] from {\x} {\y} to {\xl} {\yl}}

%%
%% Temos agora a defini{\c c}{\~a}o do comando "cmor"
%%
\def\cmor#1 #2(#3,#4)#5{%
  \!ifnextchar[{\!cmora{#1}{#2}{#3}{#4}{#5}}{\!cmora{#1}{#2}{#3}{#4}{#5}[0] }}
\def\!cmora#1#2#3#4#5[#6]{%
  \ifcase #2% para cima "\pup" (pointing up)
      \auxa=0% x mant{\^e}m-se
      \auxb=1% o y "sobe" 
    \or% para baixo "\pdown" (pointing down)
      \auxa=0% x mant{\^e}m-se
      \auxb=-1% o y "desce" 
    \or% para a direita "\pright" (pointing right)
      \auxa=1% o x move-se para a direita
      \auxb=0% o y mant{\^e}m-se
    \or% para a esquerda "\pleft" (pointing left)
      \auxa=-1% o x move-se para a esquerda
      \auxb=0% o y mant{\^e}m-se
    \fi  % constru{\c c}{\~a}o do arco
  \ifcase #6  % arco (com seta) s{\'o}lido
    \modifplot#1% Desenhar o arco
    % constru{\c c}{\~a}o da seta
    \setaxy#1
  \or  % arco (com seta) a tracejado
    \setdashes
    \modifplot#1% Desenhar o arco
    \setaxy#1
    \setsolid
  \or  % arco sem seta
    \modifplot#1% Desenhar o arco
  \fi  % seta de injec{\c c}{\~a}o
%% coloca{\c c}{\~a}o da etiqueta do morfismo
  \x=#3%  
  \y=#4%
  \multiply \x by \expansao%
  \multiply \y by \expansao%
  \put {#5} at {\x} {\y}}

%%
%% Comando para construir os Objectos
%%  \obj(x,y){<texto>}[<etiqueta>]
%% 
\def\obj(#1,#2){%
  \!ifnextchar[{\!obja{#1}{#2}}{\!obja{#1}{#2}[Nulo]}}
\def\!obja#1#2[#3]#4{%
  \!ifnextchar[{\!objb{#1}{#2}{#3}{#4}}{\!objb{#1}{#2}{#3}{#4}[1]}}
\def\!objb#1#2#3#4[#5]{%
  \x=#1%
  \y=#2%
  \def\!pinta{\normalsize$\bullet$}% para definir o tamanho normal das pintas
  \def\!nulo{Nulo}%
  \def\!arg{#3}%
  \!compara{\!arg}{\!nulo}%
  \ifcompara\def\!arg{#4}\fi%
  \multiply \x by \expansao%
  \multiply \y by \expansao%
  \setbox\caixa=\hbox{#4}%
  \!coloca{(\!arg)(#1,#2)(\number\ht\caixa,\number\wd\caixa,\number\dp\caixa)}{\pilha}%
  \auxa=\wd\caixa \divide \auxa by 131072 
  \advance \auxa by 5
  \auxb=\ht\caixa
  \advance \auxb by \number\dp\caixa
  \divide \auxb by 131072 
  \advance \auxb by 5
%(\number\auxa,
%\number\auxb)
%  \aux=\ht\caixa \divide \auxa by 131072 
% \advance \auxa by 5 
%  \auxb=\dp\caixa \divide \auxb by 131072 
%  \advance \auxb by 8
  \ifcase \tipografo    % diagramas comutativos
    \put{#4} at {\x} {\y}
  \or                   % grafos dirigidos
    \ifcase #5 % c=0
      \put{#4} at {\x} {\y}
    \or        % n=1
      \put{\!pinta} at {\x} {\y}
      \advance \y by \number\auxb  % height+depth+5
      \put{#4} at {\x} {\y}
    \or        % ne=2
      \put{\!pinta} at {\x} {\y}
      \advance \auxa by -2  % para fazer o ajuste (imperfeito)
      \advance \auxb by -2  % ao raio da circunferÍncia de centro (x,y)
      \advance \x by \number\auxa  % width+5
      \advance \y by \number\auxb  % height+depth+5
      \put{#4} at {\x} {\y}   
    \or        % e=3
      \put{\!pinta} at {\x} {\y}
      \advance \x by \number\auxa  % width+5
      \put{#4} at {\x} {\y}   
    \or        % se=4
      \put{\!pinta} at {\x} {\y}
      \advance \auxa by -2  % para fazer o ajuste (imperfeito)
      \advance \auxb by -2  % ao raio da circunferÍncia de centro (x,y)
      \advance \x by \number\auxa  % width+5
      \advance \y by -\number\auxb  % height+depth+5
      \put{#4} at {\x} {\y}   
    \or        % s=5
      \put{\!pinta} at {\x} {\y}
      \advance \y by -\number\auxb  % height+depth+5
      \put{#4} at {\x} {\y}   
    \or        % sw=6
      \put{\!pinta} at {\x} {\y}
      \advance \auxa by -2  % para fazer o ajuste (imperfeito)
      \advance \auxb by -2  % ao raio da circunferÍncia de centro (x,y)
      \advance \x by -\number\auxa  % width+5
      \advance \y by -\number\auxb  % height+depth+5
      \put{#4} at {\x} {\y}   
    \or        % w=7
      \put{\!pinta} at {\x} {\y}
      \advance \x by -\number\auxa  % width+5
      \put{#4} at {\x} {\y}   
    \or        % nw=8
      \put{\!pinta} at {\x} {\y}
      \advance \auxa by -2  % para fazer o ajuste (imperfeito)
      \advance \auxb by -2  % ao raio da circunferÍncia de centro (x,y)
      \advance \x by -\number\auxa  % width+5
      \advance \y by \number\auxb  % height+depth+5
      \put{#4} at {\x} {\y}   
    \fi
  \or                   % grafos n?o dirigidos
    \ifcase #5 % c=0
      \put{#4} at {\x} {\y}
    \or        % n=1
      \put{\!pinta} at {\x} {\y}
      \advance \y by \number\auxb  % height+depth+5
      \put{#4} at {\x} {\y}
    \or        % ne=2
      \put{\!pinta} at {\x} {\y}
      \advance \auxa by -2  % para fazer o ajuste (imperfeito)
      \advance \auxb by -2  % ao raio da circunferÍncia de centro (x,y)
      \advance \x by \number\auxa  % width+5
      \advance \y by \number\auxb  % height+depth+5
      \put{#4} at {\x} {\y}   
    \or        % e=3
      \put{\!pinta} at {\x} {\y}
      \advance \x by \number\auxa  % width+5
      \put{#4} at {\x} {\y}   
    \or        % se=4
      \put{\!pinta} at {\x} {\y}
      \advance \auxa by -2  % ver acima
      \advance \auxb by -2
      \advance \x by \number\auxa  % width+5
      \advance \y by -\number\auxb % height+depth+5
      \put{#4} at {\x} {\y}   
    \or        % s=5
      \put{\!pinta} at {\x} {\y}
      \advance \y by -\number\auxb % height+depth+5
      \put{#4} at {\x} {\y}   
    \or        % sw=6
      \put{\!pinta} at {\x} {\y}
      \advance \auxa by -2  % ver acima
      \advance \auxb by -2
      \advance \x by -\number\auxa % width+5
      \advance \y by -\number\auxb % height+depth+5
      \put{#4} at {\x} {\y}   
    \or        % w=7
      \put{\!pinta} at {\x} {\y}
      \advance \x by -\number\auxa % width+5
      \put{#4} at {\x} {\y}   
    \or        % nw=8
      \put{\!pinta} at {\x} {\y}
      \advance \auxa by -2  % ver acima
      \advance \auxb by -2
      \advance \x by -\number\auxa % width+5
      \advance \y by \number\auxb  % height+depth+5
      \put{#4} at {\x} {\y}   
    \fi
%  \or % grafos dirigidos com molduras circulares nos objectos
%    \advance \auxa by 4
%    \put{\circle{\auxa}} [Bl] at {\x} {\y}
%    \put{#4} at {\x} {\y}
%  \or % grafos n?o dirigidos com molduras circulares nos objectos
   \else % grafos com molduras circulares nos objectos
     \ifnum\auxa<\auxb % determina a maior das dimens?es
       \aux=\auxb
     \else
       \aux=\auxa
     \fi
% se a largura da caixa È menor do que 1em ent?o o tamanho 
% tamanho È ajustado para esse valor mÌnimo
     \ifdim\wd\caixa<1em
       \dimen99 = 1em
       \aux=\dimen99 \divide \aux by 131072 
       \advance \aux by 5
     \fi
     \advance\aux by -2 %folga entre o objecto e a moldura
     \multiply\aux by 2 % 
     \ifnum\aux<30
       \put{\circle{\aux}} [Bl] at {\x} {\y}
     \else
       \multiply\auxa by 2
       \multiply\auxb by 2
       \put{\oval(\auxa,\auxb)} [Bl] at {\x} {\y}
     \fi
     \put{#4} at {\x} {\y}
   \fi   
}

\catcode`!=12 %  *****  THIS MUST NEVER BE OMITTED (Ver PiCTeX)

%  \input miniltx
 % \def\Gin@driver{pdftex.def}
%  \input color.sty
 % \input graphicx.sty
 % \resetatcatcode

%\input graphicx.tex
%
%    Bundle of my macros.    Version 1.2.0.beta
%    The best use is to paste all of them into the papers
%     1/8/2005
%

%
%    Fonts.    Version 1.2.0.beta
%    The best use is to paste all of them into the papers
%     1/8/2005
%
%
% History:
%	3 Agosto 2005: ChernSimons.tex
%
%%%%%%%%%%%%%%%%%%%%%%%%%%%%

%%%%%%%%%%%%%%%%%%%%%%%%%%%%%%%%%%%%%%%%%%%%%%%%%%%%
%% Font Types	%%%%%%%%%%%%%%%%%%%%%%%%%%%%%%%%%%%%%%%%%%%%
%%%%%%%%%%%%%%%%%%%%%%%%%%%%%%%%%%%%%%%%%%%%%%%%%%%%%
\def\Serif{cmr}
\def\SerifBold{cmbx}
\def\SerifItalics{cmti}
\def\SerifSlanted{cmsl}
\def\SerifBoldItalics{cmbxti}
\def\SansSerif{cmss}
\def\SansSerifBold{cmssbx}
\def\SansSerifItalics{cmssi}
\def\SansSerifSlanted{cmssi}%%
\def\Math{cmmi}
\def\Symbols{cmsy}
\def\MathBold{cmmib}
\def\MoreSymbols{cmex}
\def\Typewriter{cmtt}
\def\Gothic{eufm}
\def\Double{msbm}
\def\Relazioni{msam}

%% Font Declarations	
%\font\tenbg=cmmib10%
%\def\bg{\tenbg}%
%%%%%%%%%%%%%%%%%%%%%%%%%%%%%%%%%%%%%%%%%%%%%%%%%%%%%%
%%%	5		%%%%%%%%%%%%%%%%%%%%%%%%%%%%%%%%%%%%%%%%%%%%
%%%	%%%%%%%%%%%%%%%%%%%%%%
= 			\Serif10 			at 5pt
= 		\SerifBold10 		at 5pt
= 	\SerifItalics10 	at 5pt
=		\SerifSlanted10 	at 5pt
=	\SerifBoldItalics10	at 5pt
= 		\SansSerif10 		at 5pt
=	\SansSerifBold10	at 5pt
=	\SansSerifItalics10	at 5pt
=	\SansSerifSlanted10	at 5pt
=				\Math10				at 5pt
=			\MathBold10			at 5pt
=			\Symbols10			at 5pt
=		\MoreSymbols10		at 5pt
=		\Typewriter10		at 5pt
=			\Gothic10			at 5pt
=			\Double10			at 5pt

%%%	7		%%%%%%%%%%%%%%%%%%%%%%%%%%%%%%%%%%%%%%%%%%%
%%%	%%%%%%%%%%%%%%%%%%%%%%%
= 			\Serif10 			at 7pt
= 		\SerifBold10 		at 7pt
= 	\SerifItalics10 	at 7pt
=	\SerifSlanted10 	at 7pt
=\SerifBoldItalics10	at 7pt
= 		\SansSerif10 		at 7pt
= 	\SansSerifBold10 	at 7pt
=\SansSerifItalics10	at 7pt
=\SansSerifSlanted10	at 7pt
=			\Math10				at 7pt
=		\MathBold10			at 7pt
=			\Symbols10			at 7pt
=		\MoreSymbols10		at 7pt
=		\Typewriter10		at 7pt
=			\Gothic10			at 7pt
=			\Double10			at 7pt

%%%	8		%%%%%%%%%%%%%%%%%%%%%%%%%%%%%%%%%%%%%%%%%
%%%	%%%%%%%%%%%%%%%%%%%%%%%%%
= 			\Serif10 			at 8pt
= 		\SerifBold10 		at 8pt
= 	\SerifItalics10 	at 8pt
=	\SerifSlanted10 	at 8pt
=\SerifBoldItalics10	at 8pt
= 		\SansSerif10 		at 8pt
= 	\SansSerifBold10 	at 8pt
=\SansSerifItalics10 at 8pt
=\SansSerifSlanted10 at 8pt
=			\Math10				at 8pt
=		\MathBold10			at 8pt
=			\Symbols10			at 8pt
=		\MoreSymbols10		at 8pt
=		\Typewriter10		at 8pt
=			\Gothic10			at 8pt
=			\Double10			at 8pt

%%%	10		%%%%%%%%%%%%%%%%%%%%%%%%%%%%%%%%%%%%%
%%%	%%%%%%%%%%%%%%%%%%%%%%%%%%%%%
= 			\Serif10 			at 10pt
= 		\SerifBold10 		at 10pt
= 		\SerifItalics10 	at 10pt
=		\SerifSlanted10 	at 10pt
=	\SerifBoldItalics10	at 10pt
= 		\SansSerif10 		at 10pt
= 	\SansSerifBold10 	at 10pt
= 	\SansSerifItalics10 at 10pt
= 	\SansSerifSlanted10 at 10pt
=				\Math10				at 10pt
=			\MathBold10			at 10pt
=			\Symbols10			at 10pt
=		\MoreSymbols10		at 10pt
=		\Typewriter10		at 10pt
=			\Gothic10			at 10pt
=			\Double10			at 10pt
=			\Relazioni10			at 10pt

%%%	12		%%%%%%%%%%%%%%%%%%%%%%%%%%%%%%%%%%%%%
%%%	%%%%%%%%%%%%%%%%%%%%%%%%%%%%%
= 				\Serif10 			at 12pt
= 			\SerifBold10 		at 12pt
= 		\SerifItalics10 	at 12pt
=		\SerifSlanted10 	at 12pt
=	\SerifBoldItalics10	at 12pt
= 			\SansSerif10 		at 12pt
= 		\SansSerifBold10 	at 12pt
= 	\SansSerifItalics10 at 12pt
= 	\SansSerifSlanted10 at 12pt
=				\Math10				at 12pt
=			\MathBold10			at 12pt
=			\Symbols10			at 12pt
=		\MoreSymbols10		at 12pt
=			\Typewriter10		at 12pt
=				\Gothic10			at 12pt
=				\Double10			at 12pt

%%%	14		%%%%%%%%%%%%%%%%%%%%%%%%%%%%%%%%
= 			\Serif10 			at 14pt
= 		\SerifBold10 		at 14pt
= 	\SerifItalics10 	at 14pt
=		\SerifSlanted10 	at 14pt
=	\SerifBoldItalics10	at 14pt
= 		\SansSerif10 		at 14pt
= 	\SansSerifBold10 	at 14pt
= \SansSerifSlanted10 at 14pt
= \SansSerifItalics10 at 14pt
=				\Math10				at 14pt
=			\MathBold10			at 14pt
=			\Symbols10			at 14pt
=		\MoreSymbols10		at 14pt
=		\Typewriter10		at 14pt
=			\Gothic10			at 14pt
=			\Double10			at 14pt

%% Styles	%%%%%%%%%%%%%%%%%%%%%%%%%%%%%%%%%%%%%%%%%%%%%
%% %%%%%%%%%%%%%%%%%%%%%%%%%%%%%%%%%%%%%%%%%%%%%
\def\NormalStyle{\parindent=5pt\parskip=3pt\normalbaselineskip=14pt%
\def\nt{\tenSerif}%
\def\rm{\fam0\tenSerif}%
\textfont0=\tenSerif\scriptfont0=\sevenSerif\scriptscriptfont0=\fiveSerif%text(\tenrm)
\textfont1=\tenMath\scriptfont1=\sevenMath\scriptscriptfont1=\fiveMath%math(\tenmi)
\textfont2=\tenSymbols\scriptfont2=\sevenSymbols\scriptscriptfont2=\fiveSymbols%symbol(\tensy)
\textfont3=\tenMoreSymbols\scriptfont3=\sevenMoreSymbols\scriptscriptfont3=\fiveMoreSymbols%ex(tenex)
\textfont\itfam=\tenSerifItalics\def\it{\fam\itfam\tenSerifItalics}%
\textfont\slfam=\tenSerifSlanted\def\sl{\fam\slfam\tenSerifSlanted}%
\textfont\ttfam=\tenTypewriter\def\tt{\fam\ttfam\tenTypewriter}%
\textfont\bffam=\tenSerifBold%
\def\bf{\fam\bffam\tenSerifBold}\scriptfont\bffam=\sevenSerifBold\scriptscriptfont\bffam=\fiveSerifBold%
\def\cal{\tenSymbols}%
\def\greekbold{\tenMathBold}%
\def\gothic{\tenGothic}%
\def\Bbb{\tenDouble}%
\def\LieFont{\tenSerifItalics}%
\nt\normalbaselines\baselineskip=14pt%
}

%%%%%%%%%%%%%%%%%%%%%%%%%%%%%%%%%%%%%%%%%%%%%%%%%%%%%%%
%%%%%%%%%%%%%%%%%%%%%%
\def\TitleStyle{\parindent=0pt\parskip=0pt\normalbaselineskip=15pt%
\def\nt{\fourteenSansSerifBold}%
\def\rm{\fam0\fourteenSansSerifBold}%
\textfont0=\fourteenSansSerifBold\scriptfont0=\tenSansSerifBold\scriptscriptfont0=\eightSansSerifBold%text(\fourteenrm)
\textfont1=\fourteenMath\scriptfont1=\tenMath\scriptscriptfont1=\eightMath%math(\fourteenmi)
\textfont2=\fourteenSymbols\scriptfont2=\tenSymbols\scriptscriptfont2=\eightSymbols%symbol(\fourteensy)
\textfont3=\fourteenMoreSymbols\scriptfont3=\tenMoreSymbols\scriptscriptfont3=\eightMoreSymbols%ex(fourteenex)
\textfont\itfam=\fourteenSansSerifItalics\def\it{\fam\itfam\fourteenSansSerifItalics}%
\textfont\slfam=\fourteenSansSerifSlanted\def\sl{\fam\slfam\fourteenSerifSansSlanted}%
\textfont\ttfam=\fourteenTypewriter\def\tt{\fam\ttfam\fourteenTypewriter}%
\textfont\bffam=\fourteenSansSerif%
\def\bf{\fam\bffam\fourteenSansSerif}\scriptfont\bffam=\tenSansSerif\scriptscriptfont\bffam=\eightSansSerif%
\def\cal{\fourteenSymbols}%
\def\greekbold{\fourteenMathBold}%
\def\gothic{\fourteenGothic}%
\def\Bbb{\fourteenDouble}%
\def\LieFont{\fourteenSerifItalics}%
\nt\normalbaselines\baselineskip=15pt%
}

%%%%%%%%%%%%%%%%%%%%%%%%%%%%%%%%%%%%%%%%%%%%%%%%%%%%%%%%
%%%%%%%%%%%%%%%%%%%%%
\def\PartStyle{\parindent=0pt\parskip=0pt\normalbaselineskip=15pt%
\def\nt{\fourteenSansSerifBold}%
\def\rm{\fam0\fourteenSansSerifBold}%
\textfont0=\fourteenSansSerifBold\scriptfont0=\tenSansSerifBold\scriptscriptfont0=\eightSansSerifBold%text(\fourteenrm)
\textfont1=\fourteenMath\scriptfont1=\tenMath\scriptscriptfont1=\eightMath%math(\fourteenmi)
\textfont2=\fourteenSymbols\scriptfont2=\tenSymbols\scriptscriptfont2=\eightSymbols%symbol(\fourteensy)
\textfont3=\fourteenMoreSymbols\scriptfont3=\tenMoreSymbols\scriptscriptfont3=\eightMoreSymbols%ex(fourteenex)
\textfont\itfam=\fourteenSansSerifItalics\def\it{\fam\itfam\fourteenSansSerifItalics}%
\textfont\slfam=\fourteenSansSerifSlanted\def\sl{\fam\slfam\fourteenSerifSansSlanted}%
\textfont\ttfam=\fourteenTypewriter\def\tt{\fam\ttfam\fourteenTypewriter}%
\textfont\bffam=\fourteenSansSerif%
\def\bf{\fam\bffam\fourteenSansSerif}\scriptfont\bffam=\tenSansSerif\scriptscriptfont\bffam=\eightSansSerif%
\def\cal{\fourteenSymbols}%
\def\greekbold{\fourteenMathBold}%
\def\gothic{\fourteenGothic}%
\def\Bbb{\fourteenDouble}%
\def\LieFont{\fourteenSerifItalics}%
\nt\normalbaselines\baselineskip=15pt%
}

%%%%%%%%%%%%%%%%%%%%%%%%%%%%%%%%%%%%%%%%%%%%%%%%%%%%%%%%
%%%%%%%%%%%%%%%%%%%%%
\def\ChapterStyle{\parindent=0pt\parskip=0pt\normalbaselineskip=15pt%
\def\nt{\fourteenSansSerifBold}%
\def\rm{\fam0\fourteenSansSerifBold}%
\textfont0=\fourteenSansSerifBold\scriptfont0=\tenSansSerifBold\scriptscriptfont0=\eightSansSerifBold%text(\fourteenrm)
\textfont1=\fourteenMath\scriptfont1=\tenMath\scriptscriptfont1=\eightMath%math(\fourteenmi)
\textfont2=\fourteenSymbols\scriptfont2=\tenSymbols\scriptscriptfont2=\eightSymbols%symbol(\fourteensy)
\textfont3=\fourteenMoreSymbols\scriptfont3=\tenMoreSymbols\scriptscriptfont3=\eightMoreSymbols%ex(fourteenex)
\textfont\itfam=\fourteenSansSerifItalics\def\it{\fam\itfam\fourteenSansSerifItalics}%
\textfont\slfam=\fourteenSansSerifSlanted\def\sl{\fam\slfam\fourteenSerifSansSlanted}%
\textfont\ttfam=\fourteenTypewriter\def\tt{\fam\ttfam\fourteenTypewriter}%
\textfont\bffam=\fourteenSansSerif%
\def\bf{\fam\bffam\fourteenSansSerif}\scriptfont\bffam=\tenSansSerif\scriptscriptfont\bffam=\eightSansSerif%
\def\cal{\fourteenSymbols}%
\def\greekbold{\fourteenMathBold}%
\def\gothic{\fourteenGothic}%
\def\Bbb{\fourteenDouble}%
\def\LieFont{\fourteenSerifItalics}%
\nt\normalbaselines\baselineskip=15pt%
}

%%%%%%%%%%%%%%%%%%%%%%%%%%%%%%%%%%%%%%%%%%%%%%%%%%%%%%%%
%%%%%%%%%%%%%%%%%%%%%
\def\SectionStyle{\parindent=0pt\parskip=0pt\normalbaselineskip=13pt%
\def\nt{\twelveSansSerifBold}%
\def\rm{\fam0\twelveSansSerifBold}%
\textfont0=\twelveSansSerifBold\scriptfont0=\eightSansSerifBold\scriptscriptfont0=\eightSansSerifBold%text(\fourteenrm)
\textfont1=\twelveMath\scriptfont1=\eightMath\scriptscriptfont1=\eightMath%math(\fourteenmi)
\textfont2=\twelveSymbols\scriptfont2=\eightSymbols\scriptscriptfont2=\eightSymbols%symbol(\fourteensy)
\textfont3=\twelveMoreSymbols\scriptfont3=\eightMoreSymbols\scriptscriptfont3=\eightMoreSymbols%ex(fourteenex)
\textfont\itfam=\twelveSansSerifItalics\def\it{\fam\itfam\twelveSansSerifItalics}%
\textfont\slfam=\twelveSansSerifSlanted\def\sl{\fam\slfam\twelveSerifSansSlanted}%
\textfont\ttfam=\twelveTypewriter\def\tt{\fam\ttfam\twelveTypewriter}%
\textfont\bffam=\twelveSansSerif%
\def\bf{\fam\bffam\twelveSansSerif}\scriptfont\bffam=\eightSansSerif\scriptscriptfont\bffam=\eightSansSerif%
\def\cal{\twelveSymbols}%
\def\bg{\twelveMathBold}%
\def\gothic{\twelveGothic}%
\def\Bbb{\twelveDouble}%
\def\LieFont{\twelveSerifItalics}%
\nt\normalbaselines\baselineskip=13pt%
}

%%%%%%%%%%%%%%%%%%%%%%%%%%%%%%%%%%%%%%%%%%%%%%%
\def\SubSectionStyle{\parindent=0pt\parskip=0pt\normalbaselineskip=13pt%
\def\nt{\twelveSansSerifItalics}%
\def\rm{\fam0\twelveSansSerifItalics}%
\textfont0=\twelveSansSerifItalics\scriptfont0=\eightSansSerifItalics\scriptscriptfont0=\eightSansSerifItalics%
\textfont1=\twelveMath\scriptfont1=\eightMath\scriptscriptfont1=\eightMath%
\textfont2=\twelveSymbols\scriptfont2=\eightSymbols\scriptscriptfont2=\eightSymbols%
\textfont3=\twelveMoreSymbols\scriptfont3=\eightMoreSymbols\scriptscriptfont3=\eightMoreSymbols%
\textfont\itfam=\twelveSansSerif\def\it{\fam\itfam\twelveSansSerif}%
\textfont\slfam=\twelveSansSerifSlanted\def\sl{\fam\slfam\twelveSerifSansSlanted}%
\textfont\ttfam=\twelveTypewriter\def\tt{\fam\ttfam\twelveTypewriter}%
\textfont\bffam=\twelveSansSerifBold%
\def\bf{\fam\bffam\twelveSansSerifBold}\scriptfont\bffam=\eightSansSerifBold\scriptscriptfont\bffam=\eightSansSerifBold%
\def\cal{\twelveSymbols}%
\def\greekbold{\twelveMathBold}%
\def\gothic{\twelveGothic}%
\def\Bbb{\twelveDouble}%
\def\LieFont{\twelveSerifItalics}%
\nt\normalbaselines\baselineskip=13pt%
}

%%%%%%%%%%%%%%%%%%%%%%%%%%%%%%%%%%%%%%%%%%%%%%%%%%%%%%%%%%%
%%%%%%%%%%%%%%%%%%
\def\AuthorStyle{\parindent=0pt\parskip=0pt\normalbaselineskip=14pt%
\def\nt{\tenSerif}%
\def\rm{\fam0\tenSerif}%
\textfont0=\tenSerif\scriptfont0=\sevenSerif\scriptscriptfont0=\fiveSerif%text(\tenrm)
\textfont1=\tenMath\scriptfont1=\sevenMath\scriptscriptfont1=\fiveMath%math(\tenmi)
\textfont2=\tenSymbols\scriptfont2=\sevenSymbols\scriptscriptfont2=\fiveSymbols%symbol(\tensy)
\textfont3=\tenMoreSymbols\scriptfont3=\sevenMoreSymbols\scriptscriptfont3=\fiveMoreSymbols%ex(tenex)
\textfont\itfam=\tenSerifItalics\def\it{\fam\itfam\tenSerifItalics}%
\textfont\slfam=\tenSerifSlanted\def\sl{\fam\slfam\tenSerifSlanted}%
\textfont\ttfam=\tenTypewriter\def\tt{\fam\ttfam\tenTypewriter}%
\textfont\bffam=\tenSerifBold%
\def\bf{\fam\bffam\tenSerifBold}\scriptfont\bffam=\sevenSerifBold\scriptscriptfont\bffam=\fiveSerifBold%
\def\cal{\tenSymbols}%
\def\greekbold{\tenMathBold}%
\def\gothic{\tenGothic}%
\def\Bbb{\tenDouble}%
\def\LieFont{\tenSerifItalics}%
\nt\normalbaselines\baselineskip=14pt%
}

%%%%%%%%%%%%%%%%%%%%%%%%%%%%%%%%%%%%%%%%%%%%%%%%%%%%%%%%%%%
%%%%%%%%%%%%%%%%%%
\def\AddressStyle{\parindent=0pt\parskip=0pt\normalbaselineskip=14pt%
\def\nt{\eightSerif}%
\def\rm{\fam0\eightSerif}%
\textfont0=\eightSerif\scriptfont0=\sevenSerif\scriptscriptfont0=\fiveSerif%text(\tenrm)
\textfont1=\eightMath\scriptfont1=\sevenMath\scriptscriptfont1=\fiveMath%math(\tenmi)
\textfont2=\eightSymbols\scriptfont2=\sevenSymbols\scriptscriptfont2=\fiveSymbols%symbol(\tensy)
\textfont3=\eightMoreSymbols\scriptfont3=\sevenMoreSymbols\scriptscriptfont3=\fiveMoreSymbols%ex(tenex)
\textfont\itfam=\eightSerifItalics\def\it{\fam\itfam\eightSerifItalics}%
\textfont\slfam=\eightSerifSlanted\def\sl{\fam\slfam\eightSerifSlanted}%
\textfont\ttfam=\eightTypewriter\def\tt{\fam\ttfam\eightTypewriter}%
\textfont\bffam=\eightSerifBold%
\def\bf{\fam\bffam\eightSerifBold}\scriptfont\bffam=\sevenSerifBold\scriptscriptfont\bffam=\fiveSerifBold%
\def\cal{\eightSymbols}%
\def\greekbold{\eightMathBold}%
\def\gothic{\eightGothic}%
\def\Bbb{\eightDouble}%
\def\LieFont{\eightSerifItalics}%
\nt\normalbaselines\baselineskip=14pt%
}

%%%%%%%%%%%%%%%%%%%%%%%%%%%%%%%%%%%%%%%%%%%%%%%%%%%%%%%%%%%
%%%%%%%%%%%%%%%%%%
\def\AbstractStyle{\parindent=0pt\parskip=0pt\normalbaselineskip=12pt%
\def\nt{\eightSerif}%
\def\rm{\fam0\eightSerif}%
\textfont0=\eightSerif\scriptfont0=\sevenSerif\scriptscriptfont0=\fiveSerif%text(\tenrm)
\textfont1=\eightMath\scriptfont1=\sevenMath\scriptscriptfont1=\fiveMath%math(\tenmi)
\textfont2=\eightSymbols\scriptfont2=\sevenSymbols\scriptscriptfont2=\fiveSymbols%symbol(\tensy)
\textfont3=\eightMoreSymbols\scriptfont3=\sevenMoreSymbols\scriptscriptfont3=\fiveMoreSymbols%ex(tenex)
\textfont\itfam=\eightSerifItalics\def\it{\fam\itfam\eightSerifItalics}%
\textfont\slfam=\eightSerifSlanted\def\sl{\fam\slfam\eightSerifSlanted}%
\textfont\ttfam=\eightTypewriter\def\tt{\fam\ttfam\eightTypewriter}%
\textfont\bffam=\eightSerifBold%
\def\bf{\fam\bffam\eightSerifBold}\scriptfont\bffam=\sevenSerifBold\scriptscriptfont\bffam=\fiveSerifBold%
\def\cal{\eightSymbols}%
\def\greekbold{\eightMathBold}%
\def\gothic{\eightGothic}%
\def\Bbb{\eightDouble}%
\def\LieFont{\eightSerifItalics}%
\nt\normalbaselines\baselineskip=12pt%
}

%%%%%%%%%%%%%%%%%%%%%%%%%%%%%%%%%%%%%%%%%%%%%
\def\RefsStyle{\parindent=0pt\parskip=0pt%
\def\nt{\eightSerif}%
\def\rm{\fam0\eightSerif}%
\textfont0=\eightSerif\scriptfont0=\sevenSerif\scriptscriptfont0=\fiveSerif%text(\tenrm)
\textfont1=\eightMath\scriptfont1=\sevenMath\scriptscriptfont1=\fiveMath%math(\tenmi)
\textfont2=\eightSymbols\scriptfont2=\sevenSymbols\scriptscriptfont2=\fiveSymbols%symbol(\tensy)
\textfont3=\eightMoreSymbols\scriptfont3=\sevenMoreSymbols\scriptscriptfont3=\fiveMoreSymbols%ex(tenex)
\textfont\itfam=\eightSerifItalics\def\it{\fam\itfam\eightSerifItalics}%
\textfont\slfam=\eightSerifSlanted\def\sl{\fam\slfam\eightSerifSlanted}%
\textfont\ttfam=\eightTypewriter\def\tt{\fam\ttfam\eightTypewriter}%
\textfont\bffam=\eightSerifBold%
\def\bf{\fam\bffam\eightSerifBold}\scriptfont\bffam=\sevenSerifBold\scriptscriptfont\bffam=\fiveSerifBold%
\def\cal{\eightSymbols}%
\def\greekbold{\eightMathBold}%
\def\gothic{\eightGothic}%
\def\Bbb{\eightDouble}%
\def\LieFont{\eightSerifItalics}%
\nt\normalbaselines\baselineskip=10pt%
}

%%%%%%%%%%%%%%%%%%%%%%%%%%%%%%%%%%%%%%%%%%%%%%%%%%%%%%%%%%%
%%%%%%%%%%%%%%%%%%%

%%%%%%%%%%%%%%%%%%%%%%%%%%%%%%%%%%%%%%%%%%%%%%%%%%%%%%%%%%%%
%%%%%%%%%%%%%%%%%%

%
%    Various Libraries.    Version 1.2.0.beta
%    The best use is to paste all of them into the papers
%     1/8/2005
%

%%%%%%%%%%%%%%%%%%%%%%%%%%%%%%
%%%%%%			Utilities		 %%%%%%
%%%%%%%%%%%%%%%%%%%%%%%%%%%%%%

% Definition modes  %
\def\ModeYes{yes}
\def\ModeNo{no}

\def\ModeUndef{undefined}

%%%%%%%%%%%%

\def\nx{\noexpand}
\def\ni{\noindent}
\def\newpage{\vfill\eject}

\def\ss{\vskip 5pt}
\def\ms{\vskip 10pt}
\def\bs{\vskip 20pt}

 \def\,{\mskip\thinmuskip}
 \def\!{\mskip-\thinmuskip}
 \def\>{\mskip\medmuskip}
 \def\;{\mskip\thickmuskip}

%%%%%%%%%%%%%%%%%%%%%%%%%%%%%%
%%%%%%		Bibliography		 %%%%%%
%%%%%%%%%%%%%%%%%%%%%%%%%%%%%%
%
% Usage:
%	[\SetModeAuto]
% ... 
%	\bib{libro1}{L.Fatibene, ...}
%	\bib{libro2}{L.Fatibene, ...}
% ...
%	(see \ref{libro2} and \ref{libro1})
% ...
% 	\ShowBiblio
%

% Definition modes  %
\def\refsModePost{post}
\def\refsModeAuto{auto}

\def\dbRefsSatusModeOk{ok}
\def\dbRefsSatusModeError{error}
\def\dbRefsSatusModeWarning{warning}

%%%%%%%%%%%%

\newcount\BNUM
\BNUM=0

\def\refs{}

\def\SetModePost{\xdef\refsMode{\refsModePost}}			%	Items are numbered by Citation order
		%	Items are numbered by Insertion order
\SetModePost

\def\dbRefsStatusOk{%
	\xdef\dbRefsStatus{\dbRefsSatusModeOk}%
	\xdef\dbRefsError{\ModeNo}%
	\xdef\dbRefsWarning{\ModeNo}%
	\xdef\dbRefsInfo{\ModeNo}%
}

\def\dbRefs{%
}

\def\dbRefsGet#1{%
	\xdef\found{N}\xdef\ikey{#1}\dbRefsStatusOk%
	\xdef\key{\ModeUndef}\xdef\tag{\ModeUndef}\xdef\tail{\ModeUndef}%
	\dbRefs%
}

\def\NextRefsTag{%
	\global\advance\BNUM by 1%
}
\def\ShowTag#1{{\bf [#1]}}

\def\dbRefsInsert#1#2{%
\dbRefsGet{#1}%
\if\found Y %
   \xdef\dbRefsStatus{\dbRefsSatusModeWarning}%
   \xdef\dbRefsWarning{record is already there}%
   \xdef\dbRefsInfo{record not inserted}%
\else%
   \toks2=\expandafter{\dbRefs}%
   \ifx\refsMode\refsModeAuto \NextRefsTag
    \xdef\dbRefs{%
   	\the\toks2 \nx\xdef\nx\dbx{#1}%
	\nx\ifx\nx\ikey %
		\nx\dbx\nx\xdef\nx\found{Y}%
		\nx\xdef\nx\key{#1}%
		\nx\xdef\nx\tag{\the\BNUM}%
		\nx\xdef\nx\tail{#2}%
	\nx\fi}%
	\global\xdef\refs{\refs \ss\ni[\the\BNUM]\ #2\par}%%%%
   \fi%   	
   \ifx\refsMode\refsModePost 
    \xdef\dbRefs{%
   	\the\toks2 \nx\xdef\nx\dbx{#1}%
	\nx\ifx\nx\ikey %
		\nx\dbx\nx\xdef\nx\found{Y}%
		\nx\xdef\nx\key{#1}%
		\nx\xdef\nx\tag{\ModeUndef}%
		\nx\xdef\nx\tail{#2}%
	\nx\fi}%
   \fi%
\fi%
}

\def\dbRefsEdit#1#2#3{\dbRefsGet{#1}%
\if\found N 
   \xdef\dbRefsStatus{\dbRefsSatusModeError}%
   \xdef\dbRefsError{record is not there}%
   \xdef\dbRefsInfo{record not edited}%
\else%
   \toks2=\expandafter{\dbRefs}%
   \xdef\dbRefs{\the\toks2%
   \nx\xdef\nx\dbx{#1}%
   \nx\ifx\nx\ikey\nx\dbx %
	\nx\xdef\nx\found{Y}%
	\nx\xdef\nx\key{#1}%
	\nx\xdef\nx\tag{#2}%
	\nx\xdef\nx\tail{#3}%
   \nx\fi}%
\fi%
}

\def\bib#1#2{\RefsStyle\dbRefsInsert{#1}{#2}%
	\ifx\dbRefsStatus\dbRefsSatusModeWarning %
		\message{^^J}%
		\message{WARNING: Reference [#1] is doubled.^^J}%
	\fi%
}

\def\ref#1{\dbRefsGet{#1}%
\ifx\found N %
  \message{^^J}%
  \message{ERROR: Reference [#1] unknown.^^J}%
  \ShowTag{??}%
\else%
	\ifx\tag\ModeUndef \NextRefsTag%
		\dbRefsEdit{#1}{\the\BNUM}{\tail}%
		\dbRefsGet{#1}%
		\global\xdef\refs{\refs \ss\ni [\tag]\ \tail\par}%%%%
	\fi
	\ShowTag{\tag}%
\fi%
}

\def\ShowBiblio{\ms\Ensure{\SectionEnsure}%
{\SectionStyle\ni References}%
{\RefsStyle\refs}%
}

%%%%%%%%%%%%%%%%%%%%%%%%%%%%%%
%%%%%%		Label DB			 %%%%%%
%%%%%%%%%%%%%%%%%%%%%%%%%%%%%%
\newcount\CHANGES
\CHANGES=0
\def\AuxFile{7}
\def\PreventDoubleOn{\xdef\PreventDoubleLabel{\ModeYes}}

\PreventDoubleOn

\def\StoreLabel#1#2{\xdef\itag{#2}% Mantiene FileAux e ritorna #2 in \itag
 \ifx\PreModeStatus\ModeNo %
   \message{^^J}%
   \errmessage{You can't use Check without starting with OpenPreMode (and finishing with ClosePreMode)^^J}%
 \else%
   \immediate\write\AuxFile{\nx\dbLabelPreInsert{#1}{\itag}}%     
   \dbLabelGet{#1}%
   \ifx\itag\tag %
   \else%
	\global\advance\CHANGES by 1%
 	\xdef\itag{(?.??)}%
    \fi%
   \fi%
}

\def\PreModeStatus{\ModeNo}

\def\ClosePreMode{\immediate\closeout\AuxFile%
  \ifnum\CHANGES=0%
	\message{^^J}%
	\message{**********************************^^J}%
	\message{**  NO CHANGES TO THE AuxFile  **^^J}%
	\message{**********************************^^J}%
 \else%
	\message{^^J}%
	\message{**************************************************^^J}%
	\message{**  PLAEASE TYPESET IT AGAIN (\the\CHANGES)  **^^J}%
    \errmessage{**************************************************^^ J}%
  \fi%
  \edef\PreModeStatus{\ModeNo}%
}

\def\dbLabelSatusModeOk{ok}

\def\dbLabelSatusModeWarning{warning}

\def\dbLabelStatusOk{%
	\xdef\dbLabelStatus{\dbLabelSatusModeOk}%
	\xdef\dbLabelError{\ModeNo}%
	\xdef\dbLabelWarning{\ModeNo}%
	\xdef\dbLabelInfo{\ModeNo}%
}

\def\dbLabel{%
}

\def\dbLabelGet#1{%
	\xdef\found{N}\xdef\ikey{#1}\dbLabelStatusOk%
	\xdef\key{\ModeUndef}\xdef\tag{\ModeUndef}\xdef\pre{\ModeUndef}%
	\dbLabel%
}

\def\ShowLabel#1{%
 \dbLabelGet{#1}%
 \ifx\tag \ModeUndef %
 	\global\advance\CHANGES by 1%
 	(?.??)%
 \else%
 	\tag%
 \fi%
}

\def\dbLabelPreInsert#1#2{\dbLabelGet{#1}%
\if\found Y %
  \xdef\dbLabelStatus{\dbLabelSatusModeWarning}%
   \xdef\dbLabelWarning{Label is already there}%
   \xdef\dbLabelInfo{Label not inserted}%
   \message{^^J}%
   \errmessage{Double pre definition of label [#1]^^J}%
\else%
   \toks2=\expandafter{\dbLabel}%
    \xdef\dbLabel{%
   	\the\toks2 \nx\xdef\nx\dbx{#1}%
	\nx\ifx\nx\ikey %
		\nx\dbx\nx\xdef\nx\found{Y}%
		\nx\xdef\nx\key{#1}%
		\nx\xdef\nx\tag{#2}%
		\nx\xdef\nx\pre{\ModeYes}%
	\nx\fi}%
\fi%
}

\def\dbLabelInsert#1#2{\dbLabelGet{#1}%
\xdef\itag{#2}%
\dbLabelGet{#1}%
\if\found Y %
	\ifx\tag\itag %
	\else%
	   \ifx\PreventDoubleLabel\ModeYes %
		\message{^^J}%
		\errmessage{Double definition of label [#1]^^J}%
	   \else%
		\message{^^J}%
		\message{Double definition of label [#1]^^J}%
	   \fi%	
	\fi%
   \xdef\dbLabelStatus{\dbLabelSatusModeWarning}%
   \xdef\dbLabelWarning{Label is already there}%
   \xdef\dbLabelInfo{Label not inserted}%
\else%
   \toks2=\expandafter{\dbLabel}%
    \xdef\dbLabel{%
   	\the\toks2 \nx\xdef\nx\dbx{#1}%
	\nx\ifx\nx\ikey %
		\nx\dbx\nx\xdef\nx\found{Y}%
		\nx\xdef\nx\key{#1}%
		\nx\xdef\nx\tag{#2}%
		\nx\xdef\nx\pre{\ModeNo}%
	\nx\fi}%
\fi%
}

%%%%%%%%%%%%%%%%%%%%%%%%%%%%%%
%%%%%%		Numbering			 %%%%%%
%%%%%%%%%%%%%%%%%%%%%%%%%%%%%%

\newcount\PART
\newcount\CHAPTER
\newcount\SECTION
\newcount\SUBSECTION
\newcount\FNUMBER
%\newdimen\TOBOTTOM
%\newdimen\LIMIT

\PART=0
\CHAPTER=0
\SECTION=0
\SUBSECTION=0	
\FNUMBER=0

\def\LastPart{\ModeUndef}
\def\LastChapter{\ModeUndef}
\def\LastSection{\ModeUndef}
\def\LastSubSection{\ModeUndef}
\def\LastClaim{\ModeUndef}
\def\Last{\ModeUndef}

\newdimen\TOBOTTOM
\newdimen\LIMIT

\def\Ensure#1{\ \par\ \immediate\LIMIT=#1\immediate\TOBOTTOM=\the\pagegoal\advance\TOBOTTOM by -\pagetotal%
\ifdim\TOBOTTOM<\LIMIT\newpage \else%
\vskip-\parskip\vskip-\parskip\vskip-\baselineskip\fi}

%%%%%%%%%%%%%%%%%%%%%%%%%%%%%
\def\PartLabel{\the\PART}
\def\NewPart#1{\global\advance\PART by 1%
         \bs\ni{\PartStyle  Part \PartLabel:}
         \bs\ni{\PartStyle #1}\newpage%
         \CHAPTER=0\SECTION=0\SUBSECTION=0\FNUMBER=0%
         \gdef\Left{#1}%
         \global\edef\Last{\PartLabel}%
         \global\edef\LastPart{\PartLabel}%
         \global\edef\LastChapter{\ModeUndef}%
         \global\edef\LastSection{\ModeUndef}%
         \global\edef\LastSubSection{\ModeUndef}%
         \global\edef\LastClaim{\ModeUndef}}
%%%%%%%%%%%%%%%%%%%%%%%%%%%%%
\def\ChapterLabel{\the\CHAPTER}
\def\NewChapter#1{\global\advance\CHAPTER by 1%
         \bs\ni{\ChapterStyle  Chapter \ChapterLabel: #1}\ms%
         \SECTION=0\SUBSECTION=0\FNUMBER=0%
         \gdef\Left{#1}%
         \global\edef\Last{\ChapterLabel}%
         \global\edef\LastChapter{\ChapterLabel}%
         \global\edef\LastSection{\ModeUndef}%
         \global\edef\LastSubSection{\ModeUndef}%
         \global\edef\LastClaim{\ModeUndef}}
%%%%%%%%%%%%%%%%%%%%%%%%%%%%%
\def\SectionEnsure{3cm}
\def\NewSection#1{\Ensure{\SectionEnsure}\gdef\SectionLabel{\the\SECTION}\global\advance\SECTION by 1%
         \ms\ni{\SectionStyle  \SectionLabel.\ #1}\ss%
         \SUBSECTION=0\FNUMBER=0%
         \gdef\Left{#1}%
         \global\edef\Last{\SectionLabel}%
         \global\edef\LastSection{\SectionLabel}%
         \global\edef\LastSubSection{\ModeUndef}%
         \global\edef\LastClaim{\ModeUndef}}
%%%%%%%%%%%%%%%%%%%%%%%%%%%%%
\def\NewAppendix#1#2{\Ensure{\SectionEnsure}\gdef\SectionLabel{#1}\global\advance\SECTION by 1%
         \bs\ni{\SectionStyle  Appendix \SectionLabel.\ #2}\ss%
         \SUBSECTION=0\FNUMBER=0%
         \gdef\Left{#2}%
         \global\edef\Last{\SectionLabel}%
         \global\edef\LastSection{\SectionLabel}%
         \global\edef\LastSubSection{\ModeUndef}%
         \global\edef\LastClaim{\ModeUndef}}
%%%%%%%%%%%%%%%%%%%%%%%%%%%%%
\def\Acknowledgements{\Ensure{\SectionEnsure}\gdef\SectionLabel{}%
         \ms\ni{\SectionStyle  Acknowledgments}\ss%
         \SECTION=0\SUBSECTION=0\FNUMBER=0%
         \gdef\Left{}%
         \global\edef\Last{\ModeUndef}%
         \global\edef\LastSection{\ModeUndef}%
         \global\edef\LastSubSection{\ModeUndef}%
         \global\edef\LastClaim{\ModeUndef}}
%%%%%%%%%%%%%%%%%%%%%%%%%%%%%
\def\SubSectionEnsure{2cm}
\def\SubSectionLabel{\ifnum\SECTION>0 \the\SECTION.\fi\the\SUBSECTION}
\def\NewSubSection#1{\Ensure{\SubSectionEnsure}\global\advance\SUBSECTION by 1%
         \ms\ni{\SubSectionStyle #1}\ss%
         \global\edef\Last{\SubSectionLabel}%
         \global\edef\LastSubSection{\SubSectionLabel}}
%%%%%%%%%%%%%%%%%%%%%%%%%%%%%
\def\SetNumberingModeN{\def\ClaimLabel{(\the\FNUMBER)}}
\def\SetNumberingModeSN{\def\ClaimLabel{(\ifnum\SECTION>0 \SectionLabel.\fi%
      \the\FNUMBER)}}
\def\SetNumberingModeCSN{\def\ClaimLabel{(\ifnum\CHAPTER>0 \the\CHAPTER.\fi%
      \ifnum\SECTION>0 \SectionLabel.\fi%
      \the\FNUMBER)}}

\def\NewClaim{\global\advance\FNUMBER by 1%
    \ClaimLabel%
    \global\edef\LastClaim{\ClaimLabel}%
    \global\edef\Last{\ClaimLabel}}
%%%%%%%%%%%%%%%%%%%%%%%%%%%%%

\def\HideLabels{\xdef\ShowLabelsMode{\ModeNo}}
\HideLabels

\def\fn{\eqno{\NewClaim}} 
\def\fl#1{%
\ifx\ShowLabelsMode\ModeYes%
%\eqno{\relax\hbox to 1cm{\NewClaim\hbox{[#1]}}}%
 \eqno{{\buildrel{\hbox{\AbstractStyle[#1]}}\over{\hfill\NewClaim}}}%
\else%
 \eqno{\NewClaim}%
\fi% 
\dbLabelInsert{#1}{\ClaimLabel}}
\def\fprel#1{\global\advance\FNUMBER by 1\StoreLabel{#1}{\ClaimLabel}%
\ifx\ShowLabelsMode\ModeYes%
%\eqno{\relax\hbox to 1cm{ .\itag\hbox{[#1]}}}%
\eqno{{\buildrel{\hbox{\AbstractStyle[#1]}}\over{\hfill.\itag}}}%
\else%
 \eqno{\itag}%
\fi% 
}

\def\cl#1{\global\advance\FNUMBER by 1\dbLabelInsert{#1}{\ClaimLabel}%
\ifx\ShowLabelsMode\ModeYes%
${\buildrel{\hbox{\AbstractStyle[#1]}}\over{\hfill\ClaimLabel}}$%
\else%
  $\ClaimLabel$%
\fi% 
}
\def\cprel#1{\global\advance\FNUMBER by 1\StoreLabel{#1}{\ClaimLabel}%
\ifx\ShowLabelsMode\ModeYes%
${\buildrel{\hbox{\AbstractStyle[#1]}}\over{\hfill.\itag}}$%
\else%
  $\itag$%
\fi% 
}
%%%%%%%%%%%%%%%%%%%%%%%%%%%%%

\def\Note{\ms\leftskip 3cm\rightskip 1.5cm\AbstractStyle}
\def\EndNote{\par\leftskip 2cm\rightskip 0cm\NormalStyle\ss}
\def\endNote{\EndNote}

%%%%%%%%%%%%%%%%%%%%%%%%%%%%%%
%%%%%%		   Sidebars		          %%%%%%
%%%%%%%%%%%%%%%%%%%%%%%%%%%%%%

\parindent=7pt
\leftskip=2cm
\newcount\SideIndent
\newcount\SideIndentTemp
\SideIndent=0
\newdimen\SectionIndent
\SectionIndent=-8pt

\def\sidebar{\vrule height15pt width.2pt }
\def\endcorner{\hbox{\hbox{\vrule height6pt width.2pt}\vbox to6pt{\vfill\hbox
to4pt{\leaders\hrule height0.2pt\hfill}}}}
\def\begincorner{\hbox{\hbox{\vrule height6pt width.2pt}\vbox to6pt{\hbox
to4pt{\leaders\hrule height0.2pt\hfill}}}}
\def\endbegincorner{\hbox{\vbox to15pt{\endcorner\vskip-6pt\begincorner\vfill}}}
\def\SideShow{\SideIndentTemp=\SideIndent \ifnum \SideIndentTemp>0 
\loop\sidebar\hskip 2pt \advance\SideIndentTemp by-1\ifnum \SideIndentTemp>1 \repeat\fi}

\def\BeginSection{{\vbadness 100000 \par\ni\hskip\SectionIndent%
\SideShow\vbox to 15pt{\vfill\begincorner}}\global\advance\SideIndent by1\vskip-10pt}

\def\EndSection{{\vbadness 100000 \par\ni\global\advance\SideIndent by-1%
\hskip\SectionIndent\SideShow\vbox to15pt{\endcorner\vfill}\vskip-10pt}}

\def\EndBeginSection{{\vbadness 100000\par\ni%
\global\advance\SideIndent by-1\hskip\SectionIndent\SideShow
\vbox to15pt{\vfill\endbegincorner}}%
\global\advance\SideIndent by1\vskip-10pt}

\def\ShowBeginCorners#1{%
\SideIndentTemp =#1 \advance\SideIndentTemp by-1%
\ifnum \SideIndentTemp>0 %
\vskip-15truept\hbox{\kern 2truept\vbox{\hbox{\begincorner}%
\ShowBeginCorners{\SideIndentTemp}\vskip-3truept}}%				
\fi%
}

\def\ShowEndCorners#1{%
\SideIndentTemp =#1 \advance\SideIndentTemp by-1%
\ifnum \SideIndentTemp>0 %
\vskip-15truept\hbox{\kern 2truept\vbox{\hbox{\endcorner}%
\ShowEndCorners{\SideIndentTemp}\vskip 2truept}}%				
\fi%
}

\def\BeginSections#1{{\vbadness 100000 \par\ni\hskip\SectionIndent%
\SideShow\vbox to 15pt{\vfill\ShowBeginCorners{#1}}}\global\advance\SideIndent by#1\vskip-10pt}

\def\EndSections#1{{\vbadness 100000 \par\ni\global\advance\SideIndent by-#1%
\hskip\SectionIndent\SideShow\vbox to15pt{\vskip15pt\ShowEndCorners{#1}\vfill}\vskip-10pt}}

\def\EndBeginSections#1#2{{\vbadness 100000\par\ni%
\global\advance\SideIndent by-#1%
\hbox{\hskip\SectionIndent\SideShow\kern-2pt%
\vbox to15pt{\vskip15pt\ShowEndCorners{#1}\vskip4pt\ShowBeginCorners{#2}}}}%
\global\advance\SideIndent by#2\vskip-10pt}

%%%%%%%%%%%%%%%%%%%%%%%%%%%%%%
%%%%%%		Margin notes		 %%%%%%
%%%%%%%%%%%%%%%%%%%%%%%%%%%%%%

%%%%%%%%%%%%%%%%%%%%%%%%%%%%%

%%%%%%%%%%%%%%%%%%%%%%%%%%%%%

%
%    Macros.    Version 1.2.0.beta
%    The best use is to paste all of them into the papers
%     1/8/2005
%

%%%%%%%%%%%%%%%%%%%%%%%%%%%%%%
%%%%%%			Greek		 %%%%%%
%%%%%%%%%%%%%%%%%%%%%%%%%%%%%%

\def\al{\alpha}
\def\be{\beta}
\def\de{\delta}
\def\ga{\gamma}

\def\ep{\epsilon}

\def\la{\lambda}

\def\om{\omega}
\def\si{\sigma}

\def\Ga{\Gamma}

%%%%%%%%%%%%%%%%%%%%%%%%%%%%%%
%%%%%%			Cal			 %%%%%%
%%%%%%%%%%%%%%%%%%%%%%%%%%%%%%

 \def\calJ{{\hbox{\cal J}}}

%%%%%%%%%%%%%%%%%%%%%%%%%%%%%%
%%%%%%			gothic		 %%%%%%
%%%%%%%%%%%%%%%%%%%%%%%%%%%%%%

 		% to prevent \sl redefinition

%%%%%%%%%%%%%%%%%%%%%%%%%%%%%%
%%%%%%			Bbb			 %%%%%%
%%%%%%%%%%%%%%%%%%%%%%%%%%%%%%

 \def\R{{\hbox{\Bbb R}}}

 \def\P{{\hbox{\Bbb P}}}
 \def\E{{\hbox{\Bbb E}}}
 \def\H{{\hbox{\Bbb H}}}
 
 \def\F{{\hbox{\Bbb F}}}

 \def\R{{\hbox{\Bbb R}}}

%%%%%%%%%%%%%%%%%%%%%%%%%%%%%%
%%%%%%		MathRoman		 %%%%%%
%%%%%%%%%%%%%%%%%%%%%%%%%%%%%%

\def\Lor{{\hbox{Lor}}}

\def\id{{\hbox{\rm id}}}

\def\cos{{\hbox{cos}}}
\def\sin{{\hbox{sin}}}

%%%%%%%%%%%%%%%%%%%%%%%%%%%%%%
%%%%%%		OtherSymbols		 %%%%%%
%%%%%%%%%%%%%%%%%%%%%%%%%%%%%%
\def\ip{\hbox to4pt{\leaders\hrule height0.3pt\hfill}\vbox to8pt{\leaders\vrule width0.3pt\vfill}\kern 2pt}
% inner product
 
\def\del{\partial}
\def\na{\nabla}

\def\Lie{\hbox{\LieFont \$}}

\def\arrLabel#1{{\buildrel {#1}\over \longrightarrow}}
\def\arr{\rightarrow}

\def\then{\Rightarrow}

%
%    Format.    Version 1.2.0.beta
%    The best use is to paste all of them into the papers
%     1/8/2005
%

%%%%%%%%%%%%%%%%%%%%
\NormalStyle
\SetNumberingModeSN
\PreventDoubleOn

\long\def\title#1{\centerline{\TitleStyle\ni#1}}

\long\def\author#1{\ms\centerline{\AuthorStyle  {\it #1}}}

\long\def\address#1{\ss\centerline{\AddressStyle #1}\par}
\long\def\moreaddress#1{\centerline{\AddressStyle #1}\par}
\def\abstract{\ms\leftskip 3cm\rightskip .5cm\AbstractStyle{\bf \ni Abstract:}\ }
\def\endabstract{\par\leftskip 2cm\rightskip 0cm\NormalStyle\ss}

%%%%%%%%%%%%%%%%%%%%%%%%%%%%%
\SetNumberingModeSN
%\ShowLabels

\def\lsim{\hbox{\lower.7ex\hbox{${\buildrel< \over \sim}$}}} 

\def\frac[#1/#2]{\hbox{$#1\over#2$}}
\def\Frac[#1/#2]{{#1\over#2}}
\def\({\left(}
\def\){\right)}
\def\[{\left[}
\def\]{\right]}
\def\^#1{{}^{#1}_{\>\cdot}}
\def\_#1{{}_{#1}^{\>\cdot}}
\def\Label=#1{{\buildrel {\hbox{\fiveSerif \ShowLabel{#1}}}\over =}}
\def\<{\kern -1pt}

%%%%%%%%  		Collapsable Notes		%%%%%%%%%%%%%%%%%%%%%%%%

\def\CollapseAllCNotes{\long\def\CNote##1{}}
\def\ExpandAllCNotes{\long\def\CNote##1{%
\BeginSection%\Margine{{\AbstractStyle To be collapsed}}%
	\Note%
 		##1%
	\endNote% 
\EndSection%
}}
\ExpandAllCNotes
%
% If you want to collapse classes of CNotes independently one of the other, just clone the definition as
%
%	\def\CollapseAllCNotesClassA{\long\def\CNoteClassA##1{}}
%	\def\ExpandAllCNotesClassA{\long\def\CNoteClassA##1{\BeginSection\Note ##1 \endNote\EndSection}}
%	\ExpandAllCNotesClassA
%
%	\def\CollapseAllCNotesClassB{\long\def\CNoteClassB##1{}}
%	\def\ExpandAllCNotesClassB{\long\def\CNoteClassB##1{\BeginSection\Note ##1 \endNote\EndSection}}
%	\ExpandAllCNotesClassB
%
%%%%%%%%%%%%%%%%%%%%%%%%%%%%%%%%%%%%%%%%%%%%%%%%%

%%%%%%%%%%%%%%%%%%%%%%%%%%%%%%%%%%%%%%%%%%%%%%%%%%%%
%
%		Links
%
%%%%%%%%%%%%%%%%%%%%%%%%%%%%%%%%%%%%%%%%%%%%%%%%%%%%
%
% \LinkFile{directory/filename}{text}
% \http{www.sigrav.org}{text}
% \Anchor{NamedDestination}
% \LinkAnchor{Name}{text}
% \NewWindow
% \NoNewWindow

\def\NoNewWindow{\edef\FlagNewWindow{/NewWindow false}}
\NoNewWindow

%

%%%%%%%%%%%%			frames 				%%%%%%%%%%%%%%%%%%%

\def\frame#1{\vbox{\hrule\hbox{\vrule\vbox{\kern2pt\hbox{\kern2pt#1\kern2pt}\kern2pt}\vrule}\hrule\kern-4pt}}

\def\Box to #1#2#3{\frame{\vtop{\hbox to #1{\hfill #2 \hfill}\hbox to #1{\hfill #3 \hfill}}}}
\def\Lag{{\bf L}}
%%%%%%%%%%%%%%%%%%%%%%%%%%%%%%%%%%%%%%%%%%%%%%%%%

%\bib{Gravitation}{}
%\bib{Ca}{}

%%%%%%%%%%%%%%%%%%%%%%%%%%%%%%%%%%%%%%%%%%%%%%%%%%%

\def\ubal{\underline{\al}\kern1pt}
\def\obal{\overline{\al}\kern1pt}

\def\ubR{\underline{R}\kern1pt}
\def\obR{\overline{R}\kern1pt}
\def\ubom{\underline{\om}\kern1pt}
\def\obxi{\overline{\xi}\kern1pt}
\def\ubu{\underline{u}\kern1pt}
\def\ube{\underline{e}\kern1pt}
\def\obe{\overline{e}\kern1pt}

%\SetModeAuto % da commentare alla fine

\bib{Casciaro1}{B.Casciaro, M.Francaviglia, V.Tapia,
{\it On the Variational Characterisation of Generalized Jacobi Equations},
Differential Geometry and Applications, Proc. Conf., Aug. 28 - Sept. 1, 1995, Brno, Czech Republic (Masaryk University, Brno, 1996), pp. 353-372;
{\tt arXiv:math-ph/0607032}
}

\bib{Casciaro2}{B. Casciaro, M. Francaviglia,
{\it Covariant second variation for first order Lagrangians on fibered manifolds. I: Generalized Jacobi fields},
Rend.~Matem.~Univ.~Roma, VII (16) (1996), pp. 233-264
}

\bib{Casciaro3}{O. Amici, B. Casciaro, M. Francaviglia,
{\it The Perturbation Functor in the Calculus of Variations},
Rend. Mat. Roma yy, xx-xx (2004)
}

\bib{Note}{  Claudio Gorodski, 
{\it An introduction to Riemannian geometry},
{\tt (to appear)}
}

\bib{Sardanashvily}{G.Sardanashvily, 
{\it Advanced Differential Geometry for Theoreticians. Fiber bundles, jet manifolds and Lagrangian theory}, 
Lambert Academic Publishing, 2013. ISBN 978-3-659-37815-7; {\tt arXiv: 0908.1886}
}

\bib{HE}{S.W.Hawking,  G.F.R. Ellis,
{\it The Large Scale Structure of Space-Time},
Cambridge University Press, 1973}

\bib{Saunders}{D.J.Saunders,
{\it The Geometry of Jet Bundles},
Cambridge University Press (1989)}

\bib{Anderson}{Ian M. Anderson,
{\it Introduction to the Variational Bicomplex},
 in Mathematical Aspects of Classical Field Theory (ed. byM. Gotay, J.Marsden, V.Moncrief), Comptemporary Mathematics Vol 132, 1992.
}

\bib{Book1}{L.Fatibene, M.Francaviglia,
{\it Natural and Gauge-Natural theories},
Kluwer, Dordrecht, (2003)}

\bib{Book2}{L.Fatibene,
{\it Relativistic theories, gravitational theories and General Relativity},
in preparation, draft version 1.0.2.
{\tt http://www.fatibene.org/book.html}
}

\bib{Palese1}{M.Francaviglia, M.Palese,
{\it Generalized Jacobi morphisms in variational sequences}, 
in: Jan Slovák and Martin Čadek (eds.): 
Proceedings of the {\it 21st Winter School "Geometry and Physics". 
Circolo Matematico di Palermo, Palermo, 2002}. 
Rendiconti del Circolo Matematico di Palermo, Serie II, Supplemento No. 69. pp. [195]—208.)}

\bib{Sternberg}{H.Goldschmidt, S. Sternberg,
{\it The Hamilton–Cartan Formalism in the Calculus of Variations}, 
Ann. Inst. Fourier, Grenoble 23 (1) (1973) 203–267.}

\bib{Palese2}{M. Francaviglia, M. Palese, R. Vitolo, 
{\it The Hessian and Jacobi Morphisms for Higher Order Calculus of Variations},
 Differential Geometry and its Applications, vol. 22 (1) (2005), p. 105-120,
ISSN: 0926-2245}

\bib{Palese3}{M. Palese, E. Winterroth, 
{\it The relation beetween the Jacobi Morphism and the Hessian in gauge-natural field theories},
 Theoretical and Mathematical Physics, vol. 152(2) (2007), p. 377-389,
ISSN: 0040-5779}

\bib{Accornero1}{L. Accornero, M. Palese,
{\it Symmetry transformations of extremals and higher conserved quantities: Invariant Yang--Mills connections},
Journal of Mathematical Physics 62, 043504 (2021)}

\bib{Accornero2}{L. Accornero, M. Palese, 
{\it The Jacobi morphism and the Hessian in higher order field theory; with applications to a Yang--Mills theory on a Minkowskian background}, 
Int. J. Geom. Meth. Mod. Phys., Vol. 17, No. 08, 2050114 (2020)}

% Il primo con Accornero lo puoi citare quando appunto studiate le proprietà di invarianza delle soluzioni (dire tipo si veda anche in un differente contesto bla bla)
% l’altro con Accornero  lo puoi citare in un contesto più generale, nell’introduzione per es., insieme a quelli con Mauro.

%%%%%%%%%%%%%%%%%%%%%%%%%%%%%%%%%%%%%%%%%%%%%%%%%%%
\NormalStyle
%\ShowLabels
\CollapseAllCNotes
\edef\PreModeStatus{\ModeYes}
	\immediate\openin\AuxFile=PreLabels.def
	\ifeof \AuxFile
	\else
 		\immediate\closeout\AuxFile
  		\input PreLabels.def
 	 \fi
	 \immediate\openout\AuxFile=PreLabels.def

\title{A variational framework for higher order perturbations}

\author{F.Chiaffredo$^1$, L.Fatibene$^{1,2}$, M.Ferraris$^1$, E.Ricossa$^1$, D.Usseglio$^{3,4}$}

\address{$^1$ Dipartimento di Matematica - University of Torino}
\moreaddress{Via C. Alberto 10, 10123, Torino, Italy}

\address{$^2$ INFN - University of Torino}
\moreaddress{Via P.Giuria, 1, 10125, Torino, Italy}

\address{$^3$ Scuola Superiore Meridionale,}
\moreaddress{ Largo San Marcellino 10, 80138, Naples, Italy}

\address{$^4$ INFN, Sezione di Napoli, Complesso Universitario di Monte S. Angelo}
\moreaddress{ Via Cintia Edificio 6, 80126, Naples, Italy}

\abstract 
a covariant, global, variational framework for perturbations in field theories is presented.
Perturbations are obtained as vertical vector fields on the configuration bundle and they drag, exactly, solution into solutions.

The flow of a perturbation drags solutions into solutions and the dragged perturbed solutions can be expanded in a series 
with respect to the flow parameter,
hence it contains perturbations at any order. 

Mechanics is included as a special case. 
As a simple application, we recover the well-known discussion about stability of geodesics on a sphere $S_2$.
\endabstract

\NewSection{Introduction}

We present a geometric framework which extends variational calculus to
(higher order) perturbations, 
aiming to  deal with perturbations and stability both in mechanics and field theory; see \ref{Casciaro1}, \ref{Casciaro2}, \ref{Casciaro3}.

We shall provide a variational derivation of generalized Jacobi equations, 
which is valid for mechanics and field theories.
Solutions of Jacobi equations are vertical vector fields on the configuration bundle, the prolongations of which are tangent to the field equations of the original Lagrangian $L$. 
Accordingly, they {\it exactly} drag solutions into solutions, at first order they reproduce first order perturbations, although they account for perturbations at any order.

Usually, first order perturbations of a solution $\bar y^i(x)$ are obtained by considering a deformation $y^i(x)\simeq  \bar y^i(x) + \de y^i(x)$ and using field equations to determine $\de y^i$
so that $y^i(x)$ are still approximate solutions {\it at first order}. 
That is more or less how one defines a vector $\de y^i$ tangent to a manifold $M$ at the point $\bar y^i$
so that the curve $\ga:t\mapsto \bar y^i + t\de y^i$ lies on the manifold {\it at first order}.
In this sense, a first order perturbation is analogous to a tangent vector to the space of solutions at a given reference solution $\bar y^i(x)$.

Jacobi fields instead are analogous to a vector field tangent to the space of solutions. Their flows preserve the space of solutions and define integral curves made of exact solutions.
This framework is then expected to provide a geometric setting to discuss, for example, gravitational waves or propagation of acoustic baryonic oscillations in cosmology.

At the same time, having a Lagrangian setting for Jacobi fields allows us to iterate the procedure and look for the analogous of second order (or higher) variations;
see \ref{Sternberg}, \ref{Palese1}, \ref{Palese2}, \ref{Palese3}, \ref{Accornero1}, \ref{Accornero2} for a different, somehow equivalent, framework. 
Second order variations are useful to discuss stability of solutions, as second derivatives are useful to discuss critical points.

The framework is based on calculus of variations on bundles to keep under control global properties and transformation laws of the objects involved.
Here, for the sake of simplicity, we shall consider first order Lagrangians even though most results extend to higher order.

The paper is organized as follows.
In Section 2, we shall briefly recall standard notation for variational calculus on fiber bundles.
In Section 3, we shall introduce composite bundles which are a slight generalization and extend variational calculus on them.

In Section 4, we introduce the Jacobi Lagrangian which determines Jacobi fields, i.e.~perturbations of solutions, of a first order Lagrangian.
Jacobi Lagrangian provides also a class of examples of variational principles on composite bundles.
We shall show in general that the  prolongation (of a suitable order) of Jacobi fields are tangent to the original field equations thought as a sub-manifold 
of a jet prolongation of the configuration bundle, in this case on $J^2B$. That also proves that a Jacobi flow drags, exactly, a solution into solutions.

In Section 5, we discuss the iteration of the process defining second variations into the variations of the variations of the original theory.
That is just as the second order tangent bundle $T^2M$ can be defined as a sub-bundle of $T(TM)$, or as the second jet prolongation $J^2B$
can be defined as a sub-bundle of $J^1(J^1B)$.
The key point is that linearization to obtain Jacobi equations can be recognized as a functor, just as $T(\cdot)$ and $J^1(\cdot)$ and, consequently, it can be iterated at will.
With second variations we can define stability and discuss it in a completely classical setting.

In Section 6, we shall discuss as an example how one can directly discuss stability of geodesics on the  sphere $S_2$; see \ref{Note}. 
That is a relatively well-known and classical result in differential geometry, although here we discuss it by a methods which closely recalls the discussion about propagation of acoustic baryonic density perturbations.

Finally, we add an Appendix to show directly that the variational framework for Jacobi equations is well suited also for second order Lagrangians.

\NewSection{Standard setting for variational calculus}

We consider a fiber bundle $[B\arrLabel{\pi} M]$, called the {\it configuration bundle}. A {\it configuration} is a (global) section $\si:M\arr B$ of the configuration bundle.
The space $\Ga(\pi)$ of all (global) sections of the configuration bundle is called the {\it configuration functional space}.

\Note
For mechanics, we can set $M=\R$. Any bundle onto a contractible base manifold, as $\R$, is trivial, thus we necessarily have $B\simeq \R\times Q$.
In this case, sections of the configuration bundle $\pi: \R\times Q \arr \R$ are in one-to-one correspondence with parameterized curves $\ga:\R\arr Q$.

More generally, we can consider $M$ to be the spacetime, so that sections are locally represented as $\si: x^\mu\mapsto (x^\mu, y^i(x))$
and $y^i(x)$ are fields on spacetime. Transition maps of the configuration bundle $[B\arr M]$ account for transformation laws of fields and one can specialize to scalar, vector, tensor fields
or any other geometric object.

In other words, the formalism can be specialized to both mechanics and field theories.
\EndNote 

In order to capture the geometric essence of variational calculus, we need  a bundle which accounts for partial derivatives of sections, just as in mechanics the tangent bundles account for the derivatives
of parameterized curves in terms of a tangent vector.
For a general introduction to jet bundles see \ref{Sardanashvily}, \ref{Saunders}, \ref{Anderson}, \ref{Book1}, \ref{Book2}.
This bundle is called the {\it (first) jet prolongation} and it is denoted by $[J^1B\arrLabel{\pi^1_0} B\arrLabel{\pi} M]$.
If $B$ has fiber coordinates $(x^\mu, y^i)$, then its first jet prolongation $J^1B$ has fibered coordinates $(x^\mu, y^i, y^i_\mu)$, where 
$y^i_\mu$ account for first partial derivatives of fields with respect to base points.

Given transition maps on $B$ in the form
$$
x'^\mu = x'^\mu(x)
\qquad
y'^i = y'^i(x, y)
\fn$$
they induce transition maps on $J^1B$ as
$$
y'^i_\mu = \bar J_\mu^\nu \( J^i_\nu + J^i_k y^k_\nu \) 
\qquad\qquad\(\hbox{where we set $\bar J_\mu^\nu=\frac[\del x^\nu/\del x'^\mu]$, $J^i_\nu=\frac[\del y'^i/\del x^\nu]$, $J^i_k=\frac[\del y'^i/\del y^k]$} \)
\fn$$

That simply means one knows how first derivatives of fields transform once one knows how fields transform.

Jet prolongations can be considered at higher orders, e.g.~$[J^2B\arrLabel{\pi^2_1} J^1B\arrLabel{\pi^1_0} B\arrLabel{\pi} M]$ is the second jet prolongation,
$J^2B$ has fiber ``coordinates'' $(x^\mu, y^i, y^i_\mu, y^i_{\mu\nu})$, where $y^i_{\mu\nu}$ account for the second order partial derivatives of fields,
they are hence understood to be symmetric in the lower indices, and they transform as second derivatives.

Jet bundles provide a framework in which systems of partial differential equations (PDE) of order $k$ become constraints and they are intrinsically 
represented as sub-manifolds $\E\subset J^{2k}B$
In the special case of mechanics, one has $J^{2k}B\simeq \R\times T^{2k}Q$ and sub-manifolds $\E\subset \R\times T^{2k}Q$ represent systems of ODE.

\Note
Let us remark that $J^k(\cdot)$ are covariant functors from bundles to bundles. In fact, $J^kB$ is a bundle, and any bundle map $\Phi:B\arr B'$ 
(which projects onto a diffeomorphism $\phi:M\arr M'$) is prolonged to a bundle morphism $J^k\Phi:J^kB\arr J^kB'$.

Accordingly, one has for free that any section $\si:M\arr B$ can be prolonged to $J^k\si:M\arr J^kB$, any fibered flow $\Phi_s$ can be prolonged to a flow $J^k\Phi_s$, any projectable vector field $\Xi$
can be prolonged to a vector field $J^k\Xi$.
That directly generalises the situation one has in mechanics with tangent bundles, tangent maps, the tangent prolongation of a curve, the prolongation of flows, 
and the tangent lift of vector fields.
\EndNote

In variational calculus, the dynamics can be given by a {\it Lagrangian}, namely a global horizontal form on the jet bundle $J^kB$.
Locally, a Lagrangian is
$$
\Lag=L(J^ky) d\si
=L(x^\mu, y^i, y^i_\mu, \dots,  y^i_{\mu_1\dots \mu_k}) d\si
\fn$$
where $d\si=dx^1\land\dots \land dx^m = d^mx$ is the local basis of $m$-forms on $M$ induced by coordinates,
hence the coefficient $L(J^ky)$ is a scalar density (iff $\Lag$ is a global form).

Since $\Lag$ is a form on $J^kB$ and $J^k\si: M\arr J^kB$, we have a form 
$$
(J^k\si)^\ast \Lag = L(x^\mu, y^i(x), \del_\mu y^i(x), \dots,  \del_{\mu_1\dots \mu_k} y^i(x)) d\si
\fn$$ on $M$ 
that can be integrated on a compact domain $D$ (with a compact boundary $\del D$)  to define the action functional
$$
A_D[\si] = \int_D (J^k\si)^\ast \Lag
\fn$$
which is guaranteed to be finite as long as fields are regular.
The action functional is a functional on the functional space  $\Ga(\pi)$ of
global sections of the configuration bundle.

A {\it deformation} on $D$ is a vertical vector field $X$ on the configuration bundle, which vanishes on the boundary $\del D$ (together with derivatives up to order $k-1$
for a Lagrangian of order $k$).
A deformation defines a vertical flow $\Phi_s:B\arr B$ on the configuration bundle, which can be prolonged to a vertical flow on the jet bundle $J^k\Phi_s:J^kB\arr J^kB$
or used to produce a 1-parameter family of configurations $\si_s=\Phi_s\circ \si$ that deforms a reference section $\si$.

We define a {\it critical configuration} $\si$ to be a section such that for all compact domains $D$ and for all deformations on $D$, the action $A_D[\si]$ is left unchanged at first
order, i.e.~such that
$$
\de_X A_D[\si] = \int_D \Frac[d/ds] (J^k\si_s)^\ast \Lag \Big|_{s=0} 
=\int_D(J^k\si)^\ast  \Frac[d/ds]  \(J^k\Phi_s\)^\ast  \Lag \Big|_{s=0} =0
\fn$$

\Note
There is a beautiful geometric way of expressing the stationarity of the action, namely, since $X$ is vertical, we obtain
$$
 \de_X\Lag := \Frac[d/ds] \(J^k\Phi_s\)^\ast  \Lag \Big|_{s=0} = \Lie_{J^kX} \Lag= i_{J^kX}  d\Lag + d\(i_{J^kX} \Lag\)= i_{J^kX}  d\Lag =0
\fn$$
which is particularly interesting because it can be written in terms of forms on a finite dimensional manifold, with no reference whatsoever to the functional space.
\EndNote
The expression we found for $\de_X \Lag=  i_{J^kX}  d\Lag $ is particularly interesting since it is a global horizontal $m$-form on $J^k V(B)$ (as a bundle over $M$) which is intrinsically given.

By a suitable series of integrations by parts, this can be recast as
$$
\de_X\Lag = \E(X) + d \(\F(J^{k-1}X)\)
\fn$$
which is called the {\it first variation formula} and it encodes everything about variational calculus in an intrinsic way.

The global horizontal $m$-form $\E(X)=\E_i X^i d\si$ is on $J^{2k}B$, it is called the {\it Euler-Lagrange form}.
The global horizontal $(m-1)$-form $\F(J^{k-1}X)$ is on $J^{2k-1}B$, it is called the {\it Liouville form}.
In the action, these forms are evaluated along the prolongation of a configuration and then integrated on $D$. In particular, one can use Stokes theorem and integrate $\F(J^{k-1}X)$ on
the boundary where the deformations $J^{k-1}X$ vanish, hence the Liouville form does not contribute to the variation of the action.
Therefore,  field equations are $\E_i=0$, while the Liouville form encodes conservation laws, which will not be discussed  here.

Any global Lagrangian $\Lag$ (of order $k$) produces global equations $\E_i = 0$. 

\Note
For a first order Lagrangian, we can set $p_i=\frac[\del L/\del y^i]$ and $p_i^\mu=\frac[\del L/\del y^i_\mu]$ and one can easily show that, in coordinates, one has
$$
\de_X\Lag = i_{J^1X}  d\Lag = \( p_iX^i +p_i^\mu X^i_\mu\) d\si
=  \( p_i -d_\mu p_i^\mu\) X^i d\si + d\( p_i^\mu X^id\si_\mu \)
\fn$$
Then we have field equations in the form  $\E_i =p_i -d_\mu p_i^\mu= 0$.
We could directly show that $\E_i'= \bar J \E_k \bar J^k_i$, hence the Euler-Lagrange form is a global form. 
\EndNote

\NewSection{Composite bundles}

Before proceeding, we wish to generalize variational principles on a bundle to composite bundles.
A {\it composite bundle} of {\it height $k$} is a ladder $[B_k\arr \dots \arr B_2\arr B_1\arr M]$ of bundles.
We have projections $\pi^{n+1}_n:B_{n+1}\arr B_n$ at each level, and compositions $\pi^{n+h}_n: B_{n+h}\arr B_n$.

A {\it composite morphism} from $[B_2\arr B_1\arr M]$ to $[B'_2\arr B'_1\arr M']$ is a triple of maps with $\Phi_2:B_2\arr B'_2$, $\Phi_1:B_1\arr B'_1$, $\Phi_0:M\arr M'$
such that they define a fibered morphism at each level (hence even at any multilevel block).  

\Note
That means that, locally, composite morphisms (as well as transition maps) are {\it stratified}, i.e.~they are in the form
$$
x'^\mu = \Phi^\mu(x)
\qquad\qquad
y'^i = \Phi^i(x, y)
\qquad\qquad
z'^a = \Phi^a(x, y, z)
\fn$$

Let us remark that a general fibered morphism, e.g.~for $[B_2\arr M]$, is not a composite morphism, just as it is not stratified.
We really restrict morphisms, hence composite bundles have additional structure to be preserved with respect to ordinary bundles.
\EndNote

A {\it composite section} of a composite bundle $[B_2\arrLabel{\pi^2_1} B_1\arrLabel{\pi^1_0} M]$ is a pair of sections $\si_1:M\arr B_1$ and $\si_2:B_1\arr B_2$.
\def\SetMargin#1{\hsize=#1\rightskip=0cm}

\Note
Locally, a composite section is given by $(y^i(x), z^a(x, y))$, while a section $\si: M\arr B_2$ is given by $(y^i(x), z^a(x))$.
Accordingly, when we define jet prolongations, we need to account for first derivatives $(y^i_\mu, z^a_\mu, z^a_i)$, as well as the second derivatives
$(y^i_{\mu\nu}, z^a_{\mu\nu}, z^a_{k\mu}, z^a_{ik})$.
\EndNote

For first derivatives, the construction is simple.
If we consider the first prolongation $J^1(\pi^1_0)$ of the bundle $[B\arr M]$, we obtain a bundle $[\pi^1_B:J^1(\pi^1_0)\arr B]$ with coordinates $(x^\mu, y^i; y^i_\mu)$.

{
{%\ms
\hfill
\rightskip -1cm
\vbox to 9cm{\begindc{\commdiag}[1]
%\obj(260, 90)[hxi]{$\hat \xi$}
%\obj(220, 50)[xi]{$\xi$}
\obj(190, 180)[P1]{$\P_1$}
\obj(190, 240)[P2]{$\P_2$}
\obj(130, 240)[JC]{$\vdots$}
\obj(130, 200)[J2C]{$J^2C$}
\obj(130, 140)[J1C]{$J^1C$}
\obj(190, 100)[C]{$C$}
\obj(190, 50)[B]{$B$}
\obj(250, 110)[J1B]{$J^1B$}
\obj(250, 170)[J2B]{$J^2B$}
\obj(250, 210)[JB]{$\vdots$}
%\obj(135, 100)[L]{$L$}
\obj(190, 0)[M]{$M$}
\obj(220, 95)[L]{$\pi^1_B$}
\mor{P1}{C}{$\pi^3_2$}
\mor{P1}{J1C}{$$}
\mor{P1}{J1B}{$$}
\mor{P2}{P1}{$\pi^4_3$}
\mor{P2}{J2C}{$$}
\mor{P2}{J2B}{$$}
\mor{B}{M}{$\pi^1_0$}
\mor{C}{B}{$\pi^2_1$}
\mor{J1B}{B}{$$}[\atright, \solidarrow]%%% un po' più in alto?
\mor{J1C}{C}{$\pi^1_C$}[\atleft, \solidarrow]%%% un po' più in alto?
\mor{J2B}{J1B}{$\pi^2_{B1}$}
\mor{JB}{J2B}{$$}
\mor{J2C}{J1C}{$\pi^2_{C1}$}
\mor{JC}{J2C}{$$}
%\mor{xi}{hxi}{}
\cmor((200,2)(208,12)(211,25)(208,45)(200,50)) \pleft(215,40){$\si$} 
\cmor((200,-3)(230,35)(248,93)) \pup(250,45){$J^1\si$} 
\cmor((180,52)(172,60)(169,76)(172,92)(180,100)) \pright(165,85){$\rho$} 
\cmor((180,50)(160,70)(130,130)) \pup(130,100){$J^1\rho$} 
%\cmor((245,95)(270,105)(278,123)) \pup(280,105){$\hat \xi$} 
\enddc
}

\SetMargin{12cm}
\vskip-9.3cm
On the other hand, we consider the first prolongation $J^1(\pi^2_1)$ of the bundle $[C\arr B]$, i.e.~a bundle $[\pi^1_C:J^1(\pi^2_1)\arr C]$ with coordinates $(x^\mu, y^i, z^a; z^a_\mu, z^a_i)$.
We  need to smash these two spaces together, which could be done by fiber product if they were bundles on the same base manifold.
Then we can define
$$
\P_1 := J^1(\pi^2_1)\times_C (\pi^2_1)^\ast \( J^1(\pi^1_0)\)
\fn$$
which, in fact, has coordinates $(x^\mu, y^i, z^a; y^i_\mu, z^a_\mu, z^a_i)$. The fiber product comes with a projection $\pi^3_2:\P_1\arr C$.

For the second jet prolongation, we consider the second prolongation $J^2(\pi^1_0)$ of the bundle $[B\arr M]$, we obtain a bundle $[\pi^2_B:J^2(\pi^1_0)\arr B]$ with coordinates $(x^\mu, y^i; y^i_\mu,  y^i_{\mu\nu})$,
as well as the second prolongation $J^2(\pi^2_1)$ of the bundle $[C\arr B]$, i.e.~a bundle $[\pi^2_C:J^2(\pi^2_1)\arr C]$ with coordinates 
$(x^\mu, y^i, z^a; z^a_\mu, z^a_i, z^a_{\mu\nu}, z^a_{i\mu}, z^a_{ij})$.

Again, we can smash these bundles together by
$$
\P_2 := J^2(\pi^2_1)\times_C (\pi^2_1)^\ast \( J^2(\pi^1_0)\)
\fn$$
which, in fact, has coordinates 

}
$$
(x^\mu, y^i, z^a; y^i_\mu, z^a_\mu, z^a_i,y^i_{\mu\nu},z^a_{\mu\nu}, z^a_{i\mu}, z^a_{ij}  )
$$
We can define a projection $\pi^4_3:\P_2\arr \P_1$.
Similarly, at higher orders.

\Note
Let us define the {\it total derivatives} 
$d_\mu y^i=y^i_\mu$ and $d_\mu z^a =  z^a_\mu + y^k_\mu z^a_k$,
as well as at second order
$d_{\mu\nu} y^i=y^i_{\mu\nu}$ and $d_{\mu\nu} z^a =  z^a_{\mu\nu} + 2z^a_{k(\mu} y^k_{\nu)} + y^k_\mu y^h_\nu z^a_{kh}+ y^k_{\mu\nu} z^a_k$.
\EndNote

Given a composite section $(\si:M\arr B, \rho:B\arr C)$
we can prolong $\si$ to $J^1\si: M\arr J^1B$ and $\rho$ to $J^1\rho: C\arr J^1C$, as in the diagram alongside.
We cannot define a section from $C$ to $\P_1$ (or from $B$ to $\P_1$).
As in the standard case, we can define, however, a section $J^1(\si,\rho): M\arr \P_1$ such that $\pi^3_2\circ J^1(\si, \rho)= \rho\circ \si$
and $\pi^3_1\circ J^1(\si, \rho)= \si$.
One starts from $x\in M$ and mark $\si(x)\in B$ and $\rho\circ\si(x)\in C$, as well as $J^1\si(x)\in J^1B$ and $J^1\rho( \si(x))\in J^1C$.

By construction, we have $\pi^1_B\(J^1\si(x)\) = \si(x) = \pi^2_1\( \rho\circ \si(x)\)$.
Now check in the definition of pull-back bundles, when one has a point $J^1\si(x)\in J^1B$ and $ \rho\circ \si(x)\in C$ such that $\pi^1_B\(J^1\si(x)\) = \pi^2_1\( \rho\circ \si(x)\)$,
that defines a point $(\rho\circ \si(x), J^1\si(x))\in (\pi^2_1)^\ast \( J^1(\pi^1_0)\)$.
The projection on $C$ is defined as $\pi^\ast: (\pi^2_1)^\ast \( J^1(\pi^1_0)\)\arr C: (\rho\circ \si(x), J^1\si(x))\mapsto \rho\circ \si(x)$.
Analogously, as one can check in the definition of product, when one has  a point $(\rho\circ \si(x), J^1\si(x))\in (\pi^2_1)^\ast \( J^1(\pi^1_0)\)$
and a point $J^1\rho\(\si(x)\)\in J^1C$
such that $\pi^\ast(\rho\circ \si(x), J^1\si(x)) = \rho\circ\si(x) = \pi^1_C(J^1\rho\(\si(x)\)) \in C$, that defines a point $\(J^1\rho\(\si(x)\), (\rho\circ \si(x), J^1\si(x))\)\in J^1C\times_C  (\pi^2_1)^\ast \( J^1(\pi^1_0)\) =\P_1$.

Then we can define a map
$$
J^1(\si,\rho): M\arr \P_1: x\mapsto \(J^1\rho\(\si(x)\), (\rho\circ \si(x), J^1\si(x))\)
\fn$$
and check it is a section by computing $\pi^3_1\circ  J^1(\si,\rho)= \si $ and
$\pi^3_2 \circ J^1(\si,\rho)=  \rho\circ \si$. 
Finally, one has $\pi^3_0 \circ J^1(\si,\rho) = \pi^1_0 \circ \si$.
In other words, $J^1(\si,\rho): M\arr \P_1$ is a section on $\rho\circ \si$ and hence  on $\si$.
Similarly, one can extend the construction to higher order prolongations.

\Note
Locally, if the composite section is expressed as $\si:M\arr B: x\mapsto (x, y(x))$ and $\rho:B\arr C: (x, y)\mapsto (x, y; z(x, y))$ (and, accordingly, 
$\rho\circ \si:M\arr C: x\mapsto (x, y(x), z(x, y(x)))$), then its prolongation to $\P_1$ is
$$
J^1(\si,\rho): M\arr \P_1:x\mapsto \(x^\mu, y^i(x), z^a(x, y(x)), \del_\mu y^i(x), \del_\mu z^a(x, y(x)), \del_i z^a(x, y(x)) \)
\fn$$ 
\EndNote

\NewSection{Linearized equations}

Let us now consider an ordinary variational principle, defined on a configuration bundle $[B\arr M]$ by a first order Lagrangian $\Lag$.
As a matter of fact, the variation of the Lagrangian $\de \Lag:J^1B \arr V^\ast(J^1 B)\otimes A_m(M)$ induces an auxiliary first order Lagrangian  
$J(\Lag)= J^1X \ip d\Lag$ 
which is called the {\it Jacobi Lagrangian}, on the 
composite bundle $[V(B)\arr B\arr M]$.
Let us stress that $J(\Lag)$ is a horizontal $m$-form in $J^1V(B)\simeq V(J^1B)$, i.e.~in the form
$$
J(\Lag)= L_J(x^\mu, y^i, y^i_\mu, X^i, X^i_\mu)\> d\si
= (p_i X^i + p_i^\mu d_\mu X^i)d\si
\fn$$  
Notice that the Jacobi Lagrangian $J(\Lag)$ happens to be linear in the vertical vector fields $X$.

Then we can consider a deformation $\Xi = \de y^i\frac[\del/\del y^i] + \de X^i \frac[\del/\del X^i] $ on $V(B)$ and the variation of the Lagrangian $J(\Lag)$.

\Note

We have
$$
\eqalign{
J^1\Xi \ip d(J(\Lag))=& (p_i \de X^i + p_i^\mu d_\mu \de X^i  ) 
+  (\del_k p_i \de y^k + \del_k^\al p_i d_\al \de y^k ) X^i  
+ (\del_k p_i^\mu \de y^k  + \del_k^\al  p_i^\mu d_\al \de y^k   ) d_\mu X^i =\cr
=& \(p_i -d_\mu p_i^\mu  \)  \de X^i+ \(  \del_k p_i  X^i + \del_k p_i^\mu  d_\mu X^i 
- d_\al\( \del_k^\al p_i  X^i + \del_k^\al  p_i^\mu  d_\mu X^i   \)  \) \de y^k +\cr
&+d_\mu\(   p_i^\mu   \de X^i +  \( \del_k^\mu p_i  X^i + \del_k^\mu  p_i^\la  d_\la X^i   \)  \de y^k  \) \cr
%=& \de^2 L 
}
\fl{JacobiEqs}$$
We obtain as a first field equation  $\E_i= p_i -d_\mu p_i^\mu =0$, i.e.~the same field equations as for the original Lagrangian $\Lag$.
For the second field equation, we can write it as
$$
X^i \del_i \E_k  +  d_\mu X^i \del_i^\mu \E_k + d_{\mu\nu} X^i \del_i^{\mu\nu} \E_k    =0
\qquad\then
J^2X (\E_i) =0
\fl{Feq}$$
which means the jet prolongation of the  vector field $X$ (vertical on the original configuration bundle $[B\arr M]$) is tangent to the original field equations
thought as a sub-manifold into $J^2B$.
\EndNote

\ni
As a consequence, the flow of $X$ drags solutions of $\Lag$ into solutions.
Because of that, the second field equation is called the {\it linearized equations} 
(or the {\it Jacobi equation}) of the original field equations $\E_k= p_k-d_\mu p_k^\mu =0$.
A solution $X$ of linearized equations is called a {\it perturbation} (or a {\it Jacobi field}).

Let us stress that the  linearized equations  come with a variational origin,
they are generated by the Jacobi Lagrangian $J(\Lag)$.

\Note
Usually, in first order perturbation theory, one  expands the original field equation at first order about a solution $\bar y^i(x)$.
That means that one considers $y^i(x)\simeq \bar y^i(x) + \ep X^i(x)$, replaces it into the field equations $\E_k=0$, keeps first order terms in $\ep$,
and interprets it as equations of $X^i$. 
When these equations are satisfied, $y^i(x)$ is still a solution {\it at first order}.

Jacobi equations are more complicated, however, the configurations generated by a flow along a Jacobi field of a solution are {\it exact} solutions.
Thus they can be expanded at any order providing us with the whole series of perturbations at any order.

To check that the two prescriptions produce the same linearized equations, we restrict, for the sake of simplicity, to a first order Lagrangian $\Lag=L(x, y, dy)d\si$
which produces equations $E_k=p_k-d_\mu p_k^\mu=0$.
The zero order term is identically satisfied since we assumed $\bar y^i$ to be a solution of the original equations.
The first order term is 
$$
X^i  \del_i p_k +  X^i_\mu\del_i^\mu p_k -d_\mu (X^i\del_i p_k^\mu    + X^i_\ep \del_i^\ep p_k^\mu) =0
\fn$$

On the other hand, when we specialize the Jacobi equations \ShowLabel{JacobiEqs} and we expand them at first order, we obtain
$$
\eqalign{
&X^i  \del_i p_k +  X^i_\mu   \del_i^\mu p_k - X^i  d_\mu \del_i p_k^\mu - X^i_\mu  \del_i^\mu d_\al p_k^\al -  X^i_{\mu\nu} \del_i^\nu  p_k^\mu =\cr
%& \qquad = X^i  \del_i p_k +  X^i_\mu   \del_i^\mu p_k - d_\mu (X^i  \del_i p_k^\mu) + \red{X^i_\mu  \del_i p_k^\mu}- `X^i_\mu d_\al \del_i^\mu  p_k^\al  - \red{X^i_\mu  \del_i  p_k^\mu}-  X^i_{\mu\nu} \del_i^\nu  p_k^\mu   =\cr
 & \qquad = X^i  \del_i p_k +  X^i_\mu   \del_i^\mu p_k - d_\mu (X^i  \del_i p_k^\mu+ X^i_\nu  \del_i^\mu  p_k^\nu) 
 =0
 }
\fn$$
Thus we see we obtain the same equations for first order perturbations, as expected.
\EndNote

Jacobi field equations \ShowLabel{Feq} are linear in $X$, therefore Jacobi fields form a real vector space $\calJ$.
Of course, that does not mean it is easy to find Jacobi fields.

\NewSection{Second variations and stability}

{\ss\hfill
%\rightskip -2cm
\vbox to 7cm{\begindc{\commdiag}[1]
%\obj(260, 90)[hxi]{$\hat \xi$}
\obj(200, 130)[JVB]{$J^1V(B)$}
\obj(220, 143)[L']{$L_J$}
\obj(230, 90)[VB]{$V(B)$}
\obj(270, 130)[VVB]{$V^2(B)$}
%\obj(220, 50)[xi]{$\xi$}
\obj(190, 50)[B]{$B$}
\obj(150, 90)[JB]{$J^1B$}
\obj(165, 100)[L]{$L$}
\obj(190, 0)[M]{$M$}
\mor{B}{M}{$\pi$}
\mor{VB}{B}{$\pi_V$}
\mor{JVB}{VB}{$\pi^1_V$}
\mor{JB}{B}{$\pi^1_0$}
\mor{VVB}{VB}{}
%\mor{xi}{hxi}{}
\cmor((200,0)(208,10)(211,25)(208,45)(200,50)) \pleft(220,25){$\si$} 
\cmor((200,55)(230,65)(238,83)) \pup(240,65){$X$} 
\cmor((245,95)(270,105)(278,123)) \pup(280,105){$\hat X$} 
\enddc
}

\vskip -7.3cm
\SetMargin{11cm}

 For linearized equations, we started on $B$, we considered a deformation $X$ on $B$, then we considered a new Lagrangian $J(\Lag)$ on $V(B)$, a new deformation $\Xi$ on $V(B)$ and consider the variation of $J(\Lag)$
 with respect to the deformation $\Xi$. Let us remark that there is no direct relation between $X$ and $\Xi$.

Now, for second order variation, we act differently. We consider a theory on $B$, consider a deformation $X= X^i\del_i$ (i.e.~a vertical vector field on $B$).
 By the flow $\Phi_s$ of $X$, we can drag a section $\si$ as $\si_s = \Phi_s \circ \si$, which is locally represented by $y^i_s(x)$ and expand it as

 }
$$
 y^i_s(x)= y^i(x) + s \de y^i(x) +\frac[1/2] s^2 \de^2 y^i(x) + O(s^3)
 \fn$$
where we set $\de y^i(x):= X^i(x, y(x))$ and $\de^2 y^i(x): = \del_j X^i(x, y(x))  \de y^j(x)$.

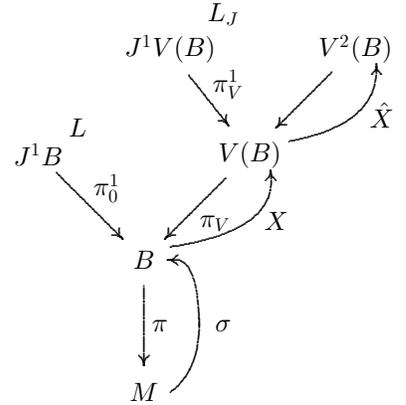
\Note
We start from a first order Lagrangian $L(x, y_s(x), d y_s(x))$. For $s=0$, we get $L$.
The first derivative is
$$
\eqalign{
&\Frac[d L/ ds]
=p_i (x, y_s, d y_s) X^i(x, y_s)+ p_i^\mu  (x, y_s, d y_s) d_\mu X^i(x, y_s)
}
\fn$$
which then specializes, for $s=0$, to $\de \Lag=J(\Lag)$.
The second derivative is then
$$
\eqalign{
&\Frac[d^2 L/ ds^2] \Big|_{s=0}\kern-10pt
= \( p_i  \>\de^2 y^i + p_i^\mu \> d_\mu  \de^2 y^i \) 
+ \(\del_k p_i \de y^k+ \del_k^\al p_i d_\al \de y^k\) \de y^i
+ \(\del_k p_i^\mu \de y^k+ \del_k^\al p_i^\mu  d_\al \de y^k\) d_\mu \de y^i \cr
}
\fn$$

Notice how that corresponds to set $X^i=\de y^i$ (and consequently $\de X^i= \de^2 y^i$) in the variation $\de\(J(\Lag)\)$ of the Jacobi Lagrangian.

\EndNote

The Lagrangian  can be expanded at second order.
$$
\eqalign{
&L(x, y_s, dy^i_{s}) = L + s \(p_i \de y^i + p_i^\mu d_\mu \de y^i\) +\frac[s^2/2]  \Big(\( p_i  \>\de^2 y^i + p_i^\mu \> d_\mu  \de^2 y^i \) +\cr
&\qquad 
+ \(\del_k p_i\> \de y^i  \de y^k + 2\del_i p_k^\al \> \de y^i d_\al \de y^k + \del_k^\al p_i^\mu \> d_\mu \de y^i  d_\al \de y^k \) \Big) +O(s^3)
}
\fn$$

To summarize, we have 
$$
L_s =L+ s \de L +\frac[1/2] s^2  \de^2 L +O(s^3)
\fn$$
where we set
$$
\eqalign{
& \de^2 \Lag  :=  \(\( p_i  \>\de^2 y^i + p_i^\mu \> d_\mu  \de^2 y^i \) 
+ \(\del_k p_i\> \de y^i  \de y^k + 2\del_i p_k^\al \> \de y^i d_\al \de y^k + \del_k^\al p_i^\mu \> d_\mu \de y^i  d_\al \de y^k \) \) d\si
}
\fn$$
Let us check that the second variation $\de^2 \Lag$ of a global first order Lagrangian $\Lag$ is a geometric object.

\Note
We consider a global Lagrangian on $B$, i.e.
$$
\Lag= L(x^\mu, y^i, y^i_\mu) d\si
\fn$$

For globality, we have $JL'(x', y'^i, y'^i_\mu)= L(x^\mu, y^i, y^i_\mu)$ and consequently
$$
p_k^\al = J \bar J^\al_\mu p'^\mu_i J^i_k
\qquad
 p_k=J p'_i J^i_k  + J \bar J^\mu_\al p'^\al_i\(J^i_{\mu k} + J^i_{jk} y^j_\mu \)= 
J \(p'_i J^i_k  + \bar J^\la_\mu p'^\mu_i (d_\la J^i_{k})\)
\fn$$

For second derivatives, we have
$$
\del_i^\mu p_k^\al =J \bar J^\al_\be \bar J^\mu_\nu \del'^\nu_jp'^\be_h J^h_k  J^j_i 
\fn$$
%$ \del_J^\mu= \bar J^\mu_\nu J_J^i \del'^\nu_i$
$$
\del_i p_k^\al = J \bar J^\al_\mu ( \del'_l  p'^\mu_j  J^j_k J_i^l  +   \bar J_\rho^\si  \del'^\rho_ l p'^\mu_j J^j_k (d_\si J^l_i )    +   p'^\mu_j J^j_{ik})
= \del_k^\al p_i
\fn$$
%$ \del_i= J_i^l \del'_l  +  \bar J_\mu^\nu (d_\nu J^l_i ) \del'^\mu_ l $

Finally, we have
$$
\eqalign{
 \del_j p_k=&J \(\del'_l p'_i J_J^l  J^i_k+ 2 \bar J_\rho^\si  \del'^\rho_ l p'_i  J^i_k   (d_\si J^l_j )+  \bar J^\la_\mu \bar J_\rho^\si\del'^\rho_ l p'^\mu_i   (d_\si J^l_j ) (d_\la J^i_{k})  + \bar J^\la_\mu p'^\mu_i (d_\la J^i_{jk})
  + p'_i J^i_{jk} \)  
  }
\fn$$
%$ \del_j= J_J^l \del'_l  +  \bar J_\rho^\si (d_\si J^l_j ) \del'^\rho_ l $

We see that, although $p_k^\al$ and $\del_i^\mu p_k^\al$ are well-defined geometrical objects, the other momenta are not.
However, there are a number of geometric objects defined out of them which should be pin pointed, since we expect to provide a good intrinsic formalism for variational calculus.

One can then show directly that $\de^2\Lag=\de^2 \Lag'$.
\EndNote

We can also expand the deformed action at second order, i.e.~as
$$
A_D[\si_s]  = \int_D (J^1 \si )^\ast \( L   +  s\>  \de L + \frac[1/2] s^2\>  \de^2 L + O(s^3) \)  
\fn$$

Suppose $\si$ is a critical section for $L$. We say that it is {\it stable} if it is a minimum of the action functional, {\it unstable} otherwise.

To decide that, we have to check how the action changes when we deform a solution $\si$. For that we need to consider the second variation.
$$
\eqalign{
 A_D[\si_s] & - A_D[\si]=  \Frac[s^2/2] \int_D  \de^2 L    + O(s^3)= \Frac[s^2/2]\( \int_{ D}  (p_i  -d_\mu p_i^\mu  ) \de^2 y^i  d\si  + \int_{\del D}   p_i^\mu   \de^2 y^i  d\si_\mu \)+ \cr
&
%\cr+&
+ \Frac[s^2/2] \int_D   \(\del_k p_i \de y^i  \de y^k + 2\del_i p_k^\al  \de y^i d_\al \de y^k + \del_k^\al p_i^\mu d_\mu \de y^i  d_\al \de y^k \) d\si +O(s^3)
\cr
}
\fn$$

Practically speaking, if the bilinear form
$$
\H =\(\matrix{\de y^i &  \de y^i_\mu }\)\(\matrix{
\del_k p_i 		&	\del_i p_k^\al	\cr
\del_k p_i^\mu	&	\del_k^\al p_i^\mu	\cr
}\)\(\matrix{
\de y^k \cr
 \de y^k_\al
}\)
\fn$$
is definite-positive, then the critical section is in a {\it minimum} of the action functional and it is called a {\it stable} solution.
Let us remark, however, that the bilinear form $\H$ is applied to a vector $(\de y^k , \de y^k_\al)$ which is holonomic, not generic.
Thus, when the form is not definite positive, we need to discuss the sign of the integral.
We shall show that in an example, in the next Section.

\NewSection{Example: unstable geodesics on a  sphere}

Let us consider, as an example, the Lagrangian for geodesics on $[B:=\R\times Q \arr \R]$ given by
$$
\Lag= \frac[1/2] g_{\mu\nu} u^\mu u^\nu ds
\fn$$
By variation, we obtain
$$
\de L= g_{\mu\nu} u^\mu \frac[d/dt]\de x^\nu +\frac[1/2] \del_\la g_{\mu\nu}u^\mu u^\nu \de x^\la 
%= u^\mu  g_{\mu\ep}\frac[d/dt]\de x^\ep +\frac[1/2] \( -\red{\del_\mu g_{\nu\la}}+ \red{\del_\nu g_{\la\mu}}+ \del_\la g_{\mu\nu} \) u^\mu u^\nu \de x^\la 
=u^\mu  g_{\mu\ep}\(  \frac[d/dt]\de x^\ep + \{g\}^\ep_{\nu\la}  u^\nu \de x^\la \)
\fn$$
Hence, the Jacobi Lagrangian on $[\R\times TQ\arr \R\times Q \arr \R]$ is
$$
J(\Lag):= \de Lds
= u^\mu g_{\mu\al}  \( \frac[d X^\al /ds]  + \{g\}^\al_{\la \nu}X^\la u^\nu \)   ds
\fn$$
which is regarded as a Lagrangian on $V(B)\simeq\R\times TQ$, which happens to be linear in $X= X^\mu(t, x)\del_\mu$
which is assumed to be a vertical vector on $\R\times Q$, or simply a vector field $X= X^\mu(x)\del_\mu$ on $Q$.
The variation of the  auxiliary Lagrangian $J(\Lag)$ defines Jacobi equation for $X$.

\Note
By variation with respect to $\de x^\mu$ and $\de X^\mu$, we obtain
$$
\eqalign{
\de (J(L))=&  
     \(  R_{\ep\mu \nu\si }X^\si   u^\mu  u^\nu 
  -   g_{\ep\be}\(  \frac[d/ds] \(\dot X^\be   + \{g\}^\be_{\si\nu}X^\si u^\nu \)   
 + \{g\}^\be_{\al\rho}  \(\dot X^\al   + \{g\}^\al_{\si\nu}X^\si u^\nu \) u^\rho \) \)   \de x^\ep +\cr
 & -    \(  \dot  u^\be     +  \{g\}^\be_{\mu\nu}   u^\mu  u^\nu   \) \( g_{\al\be} \{g\}^\al_{\si\ep}  X^\si \de x^\ep +  g_{\ep\al} \de  X^\ep \)  +\cr
&
+\frac[d /ds] \(  g_{\ep\al}  \( \dot X^\al   +  \{g\}^\al_{\si\nu}   X^\si u^\nu \) \de x^\ep
 + u^\mu g_{\mu\al}  \(  \de  X^\ep + \{g\}^\al_{\si\ep}  X^\si \de x^\ep    \)  \)
  \cr
 }
\fn$$

Since a composite section is given by $(x(t), X(t, x))$ we have
$$
\eqalign{
\dot X^\al  =& \del X^\al + u^\be d_\be X^\al
\qquad\qquad\then\quad
\dot X^\al  + \{g\}^\al_{\si\nu}X^\si u^\nu= \del X^\al + u^\be \na_\be X^\al\cr
\ddot X^\al 
%=&  \del^2 X^\al + \dot u^\be d_\be X^\al +  2u^\be d_\be \del X^\al  + u^\mu u^\nu  ( \na_{\mu\nu} X^\al  -\{g\}^\al_{\rho\mu} \na_\nu X^\rho + \{g\}^\rho_{\mu\nu}\na_\rho X^\al)- u^\mu u^\nu  d_\mu  \{g\}^\al_{\si\nu} X^\si- u^\mu u^\nu     \{g\}^\al_{\si\nu} (\na_\mu X^\si -\{g\}^\si_{\rho\mu} X^\rho)=\cr
=& \del^2 X^\al +  2u^\be \na_\be \del X^\al  
+ (\dot u^\be  + \{g\}^\be_{\mu\nu} u^\mu u^\nu) (\na_\be X^\al -\{g\}^\al_{\si\be} X^\si)-  2u^\be \{g\}^\al_{\si\be} \del X^\si  +\cr
&+ u^\mu u^\nu  ( \na_{\mu\nu} X^\al  -2\{g\}^\al_{\si\mu} \na_\nu X^\si 
%+ \{g\}^\rho_{\mu\nu}\na_\rho X^\al - \{g\}^\be_{\mu\nu} \na_\be X^\al 
- d_\mu  \{g\}^\al_{\si\nu} X^\si
+  \{g\}^\al_{\be\nu} \{g\}^\be_{\si\mu} X^\si 
+ \{g\}^\be_{\mu\nu}\{g\}^\al_{\si\be} X^\si)
%
 %u^\mu u^\nu ( - \{g\}^\be_{\mu\nu} \na_\be X^\al + \{g\}^\be_{\mu\nu}\{g\}^\al_{\si\be} X^\si )
\cr
}
\fn$$
where $\del$ denotes partial derivatives with respect to $t$. 
We also obtain along geodesics
$$
\eqalign{
  \frac[d/ds] \(\dot X^\be   + \{g\}^\be_{\si\nu}X^\si u^\nu \) 
  =  \del^2 X^\al +  2u^\be \na_\be \del X^\al  
  -  u^\be \{g\}^\al_{\si\be} \del X^\si  
+ u^\mu u^\nu  \( \na_{\mu\nu} X^\al  -\{g\}^\al_{\si\mu} \na_\nu X^\si \)
 \cr
}
\fn$$
\EndNote

Then, in view of the original equation, the Jacobi equation becomes
$$
 \del^2 X^\ep +  2u^\be \na_\be \del X^\ep  
 + u^\mu u^\nu   \na_{\mu\nu} X^\ep      
 =  u^\mu  u^\nu  R^\ep{}_{(\mu \nu)\si }X^\si  
\fn$$

\Note
This is the Jacobi equation for a time-dependent Jacobi field $X$.
In the literature, often one finds the Jacobi equations for vector fields on the manifold $Q$, i.e.~with $\del X^\al=0$, which in fact is
$$
    \na_{\mu\nu} X^\ep   = R^\ep{}_{\mu \nu\si }X^\si   
\fn$$
\EndNote

Therefore,  Euler-Lagrange equations for $J(\Lag)$, i.e.~linearized equations for $\Lag$, are
$$
\dot u^\la + \{g\}^\la_{\al\be}u^\al u^\be=0
\qquad\qquad
 \na_{\mu\nu}  X^\al  = R^\al{}_{(\nu\mu)\la}X^\la   
\fn$$
and the solution vector fields $X$ are called {\it Jacobi fieds} on $(M, g)$ and their flow sends (affinely parameterised) geodesics into  (affinely parameterised) geodesics.

The variation $\de J(\Lag)$ is something on $V(V(B))$ which can be restricted on $V^2(B)$ to get the second variation of the original Lagrangian $L$, 
evaluated along a critical configuration, and neglecting boundary term that will vanish for boundary conditions, that is
$$
\eqalign{
\de^2 L=&  %- \(u^\mu u^\nu g_{\ep\al}   \na_{\mu} \na_{\nu}  X^\al - u^\mu u^\nu R_{\ep\nu\mu\la}X^\la   \)X^\ep=
%\cr=&
  \(   g_{\ep\al}  \hat  X^\al \hat  X^\ep + u^\mu u^\nu R_{\ep\nu\mu\la}X^\la X^\ep   \) -  \frac[d/ds] \( u^\nu g_{\ep\al}   \na_{\nu}  X^\al X^\ep\)
 \cr
}
\fn$$
where we set  $\hat X^\al := \dot X^\al + u^\mu\{g\}^\al_{\mu\nu} X^\nu = u^\mu \na_\mu X^\al$.

The bilinear form $\H$ for the geodesic Lagrangian specializes to
$$
\H= \(\matrix{X^\al &  \hat X^\al }\)\(\matrix{
R_{\al\mu\nu\be} u^\mu u^\nu	&	0	\cr
0	&	g_{\al\be}	\cr
}\)\(\matrix{
X^\be\cr
\hat X^\be
}\)
\fn$$

The curvature, top-left block, on the equator of the sphere in stereographic coordinates, reads as
$$
\(\matrix{X^1 &  X^2}\)\(\matrix{
-\cos^2(t)	&	-\cos(t)\sin(t)	\cr
-\cos(t)\sin(t)	&	-\sin^2(t)	\cr
}\)\(\matrix{
X^1\cr
X^2
}\)
\fn$$
which has eigenvalues $\la=0, -1$. 
Since the metric, bottom-right block $(g_{\al\be})$ is definite positive, if we hope to get a negative value to prove instability, we need to focus on the negative eigenvector which, in this case, is $X= \cos(t)\del_x + \sin(t)\del_y$,
i.e.~it is orthogonal to the equator as one could geometrically guess.

Of course, the components $\hat X^\al$ are not independent of the component $X^\al$ (otherwise we could simply look for negative eigenvectors).
Hence, we select a vector $\xi= f(t)\( \cos(t)\del_x + \sin(t)\del_y\)$ along the equator in the direction of a negative eigenvector and $\hat \xi = \dot f(t)\( \cos(t)\del_x + \sin(t)\del_y\)$.
Then we can compute the bilinear form along $(\xi, \hat \xi)$ to be
$$
\H=  \dot f^2 - f^2
\fl{FTH}$$
We have to determine $f$ so that $\int \H \> dt<0$, which proves instability of the equator.
That depends on the boundary points; suppose we integrate between $t=0$ and $t=t_0$.
We do not need $\xi$ to be a Jacobi field, we still need it to be fixed at the boundary, $f(0)= f(t_0)=0$.

If $t_0= a \pi$, we can set
$$
f(t)= \sin\( \frac[t/a]\)
\fn$$
In fact, we have $f(0)= f(a \pi)=0$, and
$$
\int_0^{t_0} \H dt=  \frac[1/2a](1-a^2)
\fn$$
If $a<1$ (i.e.~$t_0<\pi$) that is positive and action functional in that direction is growing, which is compatible with the configuration being stable.
When $a>1$   (i.e.~$t_0>\pi$) that is negative and we found one deformation along which the action functional is not increasing which is enough to prove that the solution is unstable.

This technique is similar to consider the Fourier transform of equation \ShowLabel{FTH} and look for modes that exponentially grow.
These modes give the directions in functional space along which the action functional is changing.

\NewSection{Conclusions and perspectives}

We showed that, given any Lagrangian field theory (including, as a special case, any holonomic mechanical system)  on a bundle $[B\arr M]$,
one can define the Jacobi Lagrangian $J(\Lag) = J^kX\ip d\Lag$ on the composite bundle $[V(B)\arr B\arr M]$.
A composite configuration for the Jacobi Lagrangian is a field $y^i(x)$ and a field $X^\al(x, y)$.
That is already quite odd since usually variational principles have as configurations a number of fields on the same manifold, while composite variational principles allow to have configurations with fields defined on different manifolds.

For the Jacobi Lagrangian, $\si$ still obeys the same field equations as in the original system. 
The other field $X:B\arr V(B)$ is a vertical vector field on the configuration bundle and it obeys field equations which force it to drag solutions into solutions. 
Since the Jacobi Lagrangian is a global form, we know its equations are global or one can check it directly by changing coordinates on $[B\arr M]$, 
inducing a change of coordinates on $[V(B)\arr B\arr M]$, then on $\P_2$, computing transformation laws of momenta, and from them transformation laws of field equations, which eventually transforms as tensor densities, hence they are satisfied in one chart iff they are satisfied in any chart.

This framework applies to a variety of situations in physics.
Since the Jacobi field $X$ drags solution into solutions they can be called {\it (exact) perturbations}. 
The framework allows to discuss gravitational waves without resorting to the usual splitting $g_{\mu\nu}=\eta_{\mu\nu}+ h_{\mu\nu}$ which is nonsensical for a number of reasons, firstly because metrics are not a vector (or affine) space and linear combinations do not make sense. Moreover, the linear combination depends on the coordinates as well as on the choice of the unperturbed state $\eta$. 
Secondly, when first order Einstein equations are found, solved for a first order perturbation $h$, the corresponding $g$ is not a solution of Einstein equations, it is just a solution at first order.

The framework we presented here, instead, is the natural setting to discuss the issue in a covariant way.

The same argument holds for computing the evolution of acoustic perturbations in a cosmological setting, which is routinely used to compute the power spectrum of the cosmic microwave background radiation.

One can also consider applications to classical mathematical GR. 
Many issues with gravity directly stem from the fact that the space of fields are not linear, one for all, there is no superposition principle in GR.
Well, one can try and parameterize configurations in terms of $X$ which are (vertical) vector fields!

\Note
Proceeding backwards, one can consider a family of exact solutions, e.g.~Schwarzschild solutions which are parameterized by a parameter $m$, and 
we can define a Jacobi field (along the family) by taking the derivative with respect to the family parameter, obtaining in this way precisely a vertical vector field $X$ 
(along the family)  on the bundle $[\Lor(M)\arr M]$.
Moreover, the flow of such a Jacobi field is known, by construction, to be the family of Schwarzschild metrics themselves. 
In other words,  $X$ drags Minkowski spacetime into the whole family of Schwarzschild metrics.
\EndNote

Once we have two Jacobi fields $X_1, X_2$, we can safely sum them being linear solutions of linear equations.
Of course, there is no simple relation among the flows of $X_1, X_2, X_1+X_2$. 
The operation of taking the flow is not linear itself.
However, $ X_1+X_2$ drags Minkowski into solutions which are not the original metrics.
For example, if we consider the fields that generate the Schwarzschild family at two different points, then
$X_1+X_2$ is not spherically symmetric, it is rather what we should call the field generated by two masses.
Analytic expressions for it are probably hopeless in most cases, that does not mean one cannot 
infer properties of solutions from the fact that the geometry is produced by the flow of a particular field. 
Further investigations will be devoted in this direction.

The approach in terms of flows is not new (see \ref{HE}). 
The new part of the framework is the variational setting to compute the Jacobi equations.

Finally, let us mention that Jacobi setting also defines second variation of the original Lagrangian and define a global intrinsic notion of stability.
For example, by applying this framework one is able to prove 
Breitenlohner-Freedman bound for the mass of Klein-Gordon fields which makes the vacuum state stable in curved different backgrounds.
We shall show that in a forthcoming paper.

Thanks to the functoriality of linearization, one can iterate the variation to any order.
Indeed, for example, we have 
$$
\eqalign{
\de y^i =& X^i(x, y_s(x))\Big|_{s=0}\cr
\de^2 y^i =& \del_k  X^i(x, y_s(x)) X^k(x, y_s(x)) \Big|_{s=0} \kern-10pt=  X^i_k \de y^k\cr
\de^3 y^i =&  \del_{kn}  X^i(x, y_s(x)) X^k(x, y_s(x))X^n(x, y_s(x))  \Big|_{s=0} \kern-10pt+\del_k  X^i_k \de^2 y^k=\cr
=& X^i_k\> \de^2 y^k + X^i_{kn}\> \de y^k \de y^n \cr
\dots
}
\fn$$
at third order.

\NewAppendix{A}{Linearized equations for a general second order Lagrangians}

Let us now consider an ordinary variational principle, defined on a configuration bundle $[B\arr M]$ and described by a second order Lagrangian $\Lag=L(J^2y)d\si$.
As a matter of fact, the variation of the Lagrangian $\de \Lag:J^2B \arr V^\ast(J^2 B)\otimes A_m(M)$ is a global form and it induces an auxiliary global second order Jacobi Lagrangian  
$J(\Lag)= J^2X \ip d\Lag $ 
on the 
composite bundle $[V(B)\arr B\arr M]$.
Let us stress that $J(\Lag)$ is a horizontal $m$-form in $J^2V(B)\simeq V(J^2B)$, i.e.
$$
J(\Lag)= L_J(x, y, X, dy, dX, d^2y, d^2X)\> d\si= (p_i X^i + p_i^\mu d_\mu X^i+ p_i^{\mu\nu} d_{\mu\nu} X^i )d\si
\fn$$  

Then we can consider a deformation $\Xi = \de y^i\frac[\del/\del y^i] + \de X^i \frac[\del/\del X^i] $ on $V(B)$ and the variation of the Lagrangian $J(\Lag)$.

\Note

We have
$$
\eqalign{
J^2\Xi& \ip d(J(\Lag))= \(p_i -d_\mu p_i^\mu + d_{\mu\nu} p_i^{\mu\nu} \)  \de X^i
+ \Big(  \del_k p_i  X^i + \del_k p_i^\mu  d_\mu X^i + \del_k p_i^{\mu\nu}  d_{\mu\nu} X^i +\cr
&- d_\al\( \del_k^\al p_i  X^i + \del_k^\al  p_i^\mu  d_\mu X^i   + \del_k^\al  p_i^{\mu\nu} d_{\mu\nu} X^i \)
+d_{\al\be} \( \del_k^{\al\be} p_i  X^i + \del_k^{\al\be}  p_i^\mu  d_\mu X^i + \del_k^{\al\be}  p_i^{\mu\nu}  d_{\mu\nu} X^i  \)  \Big) \de y^k +\cr
&+d_\mu\Big(   \(p_i^\mu   -  d_{\nu} p_i^{\mu\nu} \) \de X^i + p_i^{\mu\nu} d_{\nu} \de X^i 
+  \( \del_k^\mu p_i  X^i + \del_k^\mu  p_i^\la  d_\la X^i   + \del_k^\mu  p_i^{\la\nu} d_{\la\nu} X^i \)  \de y^k+\cr
&-d_\al \( \del_k^{\al\mu} p_i  X^i + \del_k^{\al\mu}  p_i^\la  d_\la X^i + \del_k^{\al\mu}  p_i^{\la\nu}  d_{\la\nu} X^i  \) \de y^k  
+ \( \del_k^{\mu\be} p_i  X^i + \del_k^{\mu\be}  p_i^\la  d_\la X^i + \del_k^{\mu\be}  p_i^{\la\nu}  d_{\la\nu} X^i  \)  d_{\be} \de y^k
\Big) \cr
%=& \de^2 L 
}
\fn$$
We obtain as a first field equation  $\E_i= p_i -d_\mu p_i^\mu + d_{\mu\nu} p_i^{\mu\nu}=0$, i.e.~the same field equations as for the original Lagrangian $\Lag$.
For the second field equation we can write it as
$$
X^i \del_i \E_k  +  d_\mu X^i \del_i^\mu \E_k + d_{\mu\nu} X^i \del_i^{\mu\nu} \E_k
 +  d_{\mu\nu\al} X^i  \(  -  \del_i^{\mu\nu}  p_k^\al    +  2   \del_i^{\mu\nu} d_\be p_k^{\al\be}  -   \del_i^{\mu} p_k^{\al\nu}    \)
 +  d_{\mu\nu\al \be} X^i \del_i^{\mu\nu}  p_k^{\al\be}  
    =0
\fn$$
where we used that $[d_\al, \del_k] =0$, $[d_\al, \del_k^\mu] = -  \de^\mu_\al \del_k$,  $[d_\al, \del_k^{\mu\nu}]  = -  \de^{(\nu}_\al \del_k^{\mu)}$,
and $-[d_\al, \del_k^{\mu\nu\ep}]=   \del_k^{\mu\nu\ep} d_\al = \de^{(\ep}_\al \del_k^{\mu\nu)}$.
In order to further simplify the second field equation, let is compute on a side
$$
\eqalign{
d_{\al\be\ep\la}X^i \del_i^{\al\be\ep\la} \E_k=& d_{\al\be\ep\la}X^i \del_i^{\al\be\ep\la} (p_k -d_\mu p_k^\mu + d_{\mu\nu}p_k^{\mu\nu})
= d_{\al\be\ep\la}X^i \del_i^{\al\be\ep\la}  d_{\mu\nu}p_k^{\mu\nu}= d_{\al\be\ep\la}X^i   \del_i^{\al\be} p_k^{\ep\la}\cr
d_{\al\be\ep}X^i \del_i^{\al\be\ep} \E_k=& d_{\al\be\ep}X^i \del_i^{\al\be\ep} (p_k -d_\mu p_k^\mu + d_{\mu\nu}p_k^{\mu\nu})
= d_{\al\be\ep}X^i  \( -\del_i^{\be\ep}  p_k^\al  - \del_i^{\al}   p_k^{\be\ep}   + 2\del_i^{\al\be} d_\mu p_k^{\mu\ep}  \)
}
\fn$$
where we used the identities
$$
d_{\al\be\ep}X^i  \del_i^{\al\be\ep} d_{\mu\nu} p_k^{\mu\nu}=  d_{\al\be\ep}X^i (  - \del_i^{\al}   p_k^{\be\ep}   + 2\del_i^{\al\be} d_\mu p_k^{\mu\ep}) 
\fn$$
$$
d_{\al\be\ep\la}X^i  \del_i^{\al\be\ep\la}  d_{\mu\nu}p_k^{\mu\nu}=d_{\al\be\ep\la}X^i   \del_i^{\al\be} p_k^{\ep\la}
\fn$$

In view of these results, the second field equation is recasted as $J^4X (\E_i) =0$
which means the jet prolongation of the  vector field $X$ (vertical on the original configuration bundle $[B\arr M]$) is tangent to the original field equations
thought as a sub-manifold into $J^4B$.

\Acknowledgements
We also acknowledge the contribution of INFN (Iniziativa Specifica QGSKY), the local research project {\it  Metodi Geometrici in Fisica Matematica e Applicazioni (2022)} of Dipartimento di Matematica of University of Torino (Italy). This paper is also supported by INdAM-GNFM.

\ShowBiblio
\ClosePreMode

\end